\documentclass[journal]{IEEEtran}
\usepackage{cite}
\usepackage[normalem]{ulem}
\usepackage{amsmath,amssymb,amsfonts}
\usepackage{algorithmic}
\usepackage{graphicx}
\usepackage{textcomp}
\usepackage{xcolor}
\usepackage{hhline}
\usepackage{multirow}
\usepackage{anyfontsize}
\usepackage{lineno}
\usepackage[flushleft]{threeparttable}
\ifCLASSINFOpdf
\else
\fi
\ifCLASSOPTIONcompsoc
 \usepackage[caption=false,font=normalsize,labelfont=sf,textfont=sf]{subfig}
\else
 \usepackage[caption=false,font=footnotesize]{subfig}
\fi
\begin{document}
%
\title{ChipQA: No-Reference Video Quality Prediction via Space-Time Chips}
%
%
%

\author{Joshua P. Ebenezer,
        Zaixi Shang,~\IEEEmembership{Student Member,~IEEE,}
        Yongjun Wu, Hai Wei, Sriram Sethuraman, Alan C. Bovik~\IEEEmembership{Fellow,~IEEE} 
\thanks{J.P. Ebenezer, Z. Shang, and A.C. Bovik are with the Laboratory for Image and Video Engineering, The University of Texas at Austin, Austin,
TX, 78712, USA e-mail: joshuaebenezer@utexas.edu.}
\thanks{Y. Wu, H. Wei, and  S. Sethuraman are with Amazon Prime Video.}
\thanks{J.P. Ebenezer and Z. Shang contributed equally to this work}}

\maketitle

\begin{abstract}
We propose a new model for no-reference video quality assessment (VQA). Our approach uses a new idea of highly-localized space-time (ST) slices called Space-Time Chips (ST Chips). ST Chips are localized cuts of video data along directions that \textit{implicitly} capture motion. We use perceptually-motivated bandpass and normalization models to first process the video data, and then select oriented ST Chips based on how closely they fit parametric models of natural video statistics. We show that the parameters that describe these statistics can be used to reliably predict the quality of videos, without the need for a reference video. The proposed method implicitly models ST video naturalness, and deviations from naturalness. We train and test our model on several large VQA databases, and show that our model achieves state-of-the-art performance at reduced cost, without requiring motion computation. 
\end{abstract}


%
\IEEEpeerreviewmaketitle

\section{Introduction}
%
%
%
%
\IEEEPARstart{V}{ideo} content continues to proliferate, already accounting for more than 70\% of internet traffic, and projected to exceed 82\% of internet traffic by 2021. Live internet video will account for 13 percent of Internet video traffic by 2021, and is predicted to grow 15-fold from 2016 to 2021~\cite{cisco}.  Distortions can affect videos as they are captured, transmitted, and received. The task of assessing the quality of a video in the presence of distortions is thus an increasingly important open problem. In most instances in this process there is no reference against which to measure their eventual perceived quality. Nevertheless, it is of vital importance to providers of video content to be able to monitor and predict the perceptual quality of their videos, since this directly impacts customer satisfaction. Video quality tools can also help make well-informed design choices while creating systems for capturing, processing, transmitting, and displaying videos. Video quality assessment algorithms also have applications in video denoising, designing loss functions for deep learning, video compression, and many other high-impact areas.

\begin{figure}[!h] 
  \includegraphics[width=\linewidth]{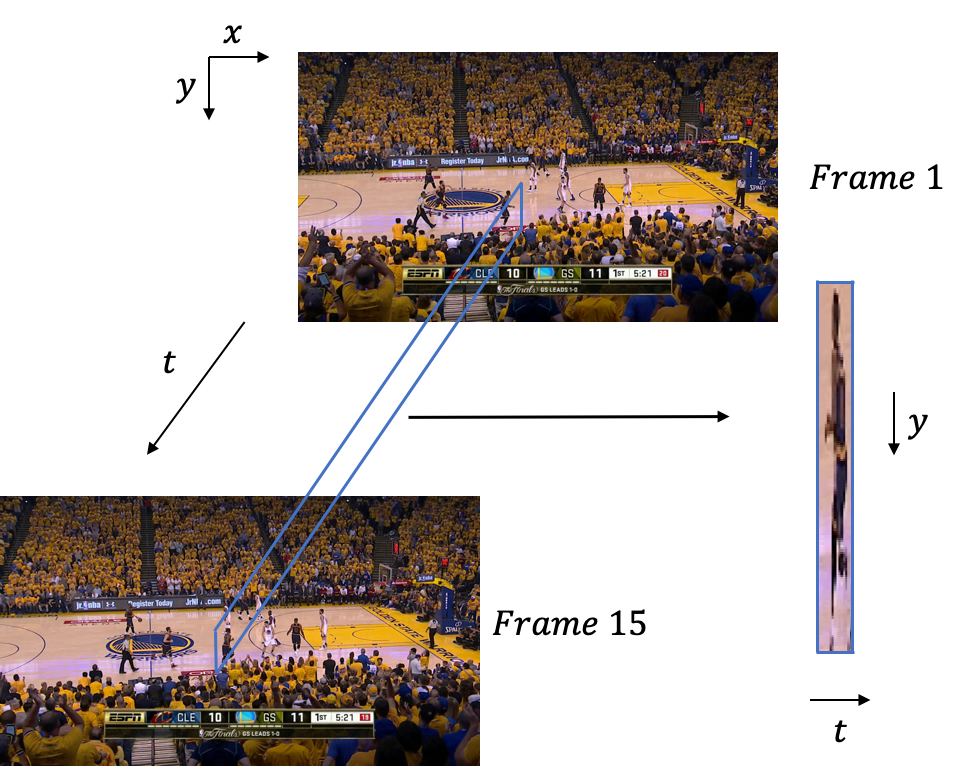}
  \caption{Space-Time Chips capture elements in motion. On the left are frames 1 and 15 of a video of a basketball game. The ST-chip is marked in blue. The player near the edges of the chip on the $xy$ plane is moving to his right as time progresses. The chip is a localized cut of all the frames between 1 and 15, perpendicular to the direction of his motion, which captures the player, 
  as shown on the right.}\label{fig:chip_illus}
\vspace{-3mm}
\end{figure}

Collecting a large number of human opinion scores on the quality of a video is the most reliable way to measure its quality. However, collecting subjective opinions of video quality is a cumbersome and expensive task. It is also time-consuming and cannot be deployed prior to or during transmission of a video, when they are being live-streamed or have other latency constraints. Subjective opinions are nevertheless useful as a gold standard when designing objective video quality assessment (VQA) algorithms. Objective VQA algorithms are designed to correlate well with these subjective human judgments, and deployed effectively and cheaply in video processing systems. VQA algorithms are typically evaluated on the basis of data gathered from studies on human judgments of video quality. Subjective judgments of video quality are first collected from a statistically significant number of human observers and normalized with respect to each observer's scores to form an opinion score for each observer and for each video. These opinion scores are then averaged across the observers yielding single mean opinion score (MOS) for each video. These mean opinion scores are the ground truth against which objective VQA algorithms are trained and tested. 

Objective VQA algorithms fall into three categories: full-reference (FR), reduced-reference (RR), and no-reference (NR). FR VQA algorithms require a reference video against which the distorted video is compared. RR VQA algorithms require only some information from the reference video, but not all, to predict the quality of a distorted video. NR VQA algorithms do not make use of a reference video, and the models we present here fall in this category. NR VQA algorithms rely on distortion-specific features or models of natural video statistics to predict video quality, and are of great interest because of their potentially broader applications.

In this work, we propose a NR VQA algorithm based on the natural video statistics of space-time (ST) chips. ST Chips are a new feature space that are defined as localized and oriented cuts of a video volume, and an illustration of the concept is shown in Fig.~\ref{fig:chip_illus}. We show that when a pristine video is processed using models derived from the human visual system, ST Chips extracted from the processed video that are along the direction of motion follow certain regular statistics, which is a breakthrough in our understanding of natural video statistics. We first proposed the idea of using ST Chips in \cite{chipqa0}, where we extracted ST Chips using optical flow in a prototype algorithm. In this work, we develop that idea further, introducing temporal processing of the video data based on models of the human visual system, and doing away with optical flow by using a simpler and more elegant approach based on regularities revealed by analysis of video statistics. Directions of motions are found in an implicit manner by using well-known models of natural image statistics and the smoothness of motion fields. The statistics of ST Chips extracted along these directions of motion can be modelled with parametrized distributions, and we show that these parameters can be used to reliably predict the quality of videos. We call our model ChipQA, which we designed to be able to handle different kinds of videos. We show that ChipQA achieves state-of-the-art (SOTA) performance on a large new high-motion VQA database. We also test ChipQA on several other VQA databases of professional and user generated content and show that it achieves high-correlations with human judgments of video quality, while also being very computationally efficient. 

The paper is organized as follows. In the following section, we briefly review previous work in the area of NR VQA. In section III, we describe our algorithm and its perceptual underpinnings. In section IV we report and analyze results on several large VQA database, and we conclude in section V.

\section{Previous Work}

V-BLIINDS \cite{vbliinds} is a no-reference video quality algorithm that models the natural video statistics of the discrete cosine transform (DCT) of frame differences. V-BLIINDS also makes use of features that capture global and local motion coherency. VIIDEO \cite{viideo} is a "completely blind" NR VQA algorithm, in that it is not trained on a database at all and can be deployed as-is. VIIDEO makes use of the high inter-subband correlations of statistical features that have been observed in pristine videos but not in distorted videos. VIIDEO predicts the quality of videos based on this observation and without any training. Manasa and Channappayya \cite{nrof} proposed an NR VQA algorithm based on the statistics of optical flow. The coefficient of variation of the standard deviation of optical flow at different spatial locations is used to quantify irregularities in motion. Dendi and Channappayya \cite{3dmscn} proposed a statistical model for the distribution of spatio-temporal bandpass coefficients. The statistics of these coefficients are modelled as following an assymmetric generalized Gaussian distribution. The parameters from the statistical fits are used to predict quality. ChipQA-0~\cite{chipqa0} introduced the idea of localized cuts in space-time, ST Chips, which may be viewed as highly localized variations of space-time slices, which are defined over the global range of spatial and temporal video coordinates, instead of locally. The ST Chips in ChipQA-0 were extracted using optical flow, making the algorithm expensive and impractical for use when low-latency is a requirement. The statistics of ST Chips are modelled based on the general observation that natural videos follow regular statistics, and that the regularity of these statistics is disturbed in the presence of distortions. Quantifying these deviations from natural statistics can thus be used to quantify the degree of distortion and the perceptual quality of the video, by learning mappings between these statistics and perception. Finding and describing these statistics is a challenge but many clues about these patterns can be gleaned from the human visual system. The human visual system has adapted to the regular statistics of videos, using them to reduce redundancies in visual signals. Mimicking the front-end visual processes involved in encoding the visual signal, it is possible to reveal departures from these regularities and use them to quantify video quality.

\begin{figure*}
\centering
\subfloat[$k(t)$ for $t \in (0,10)$ for $a=0.5$]{\includegraphics[width=2.5in]{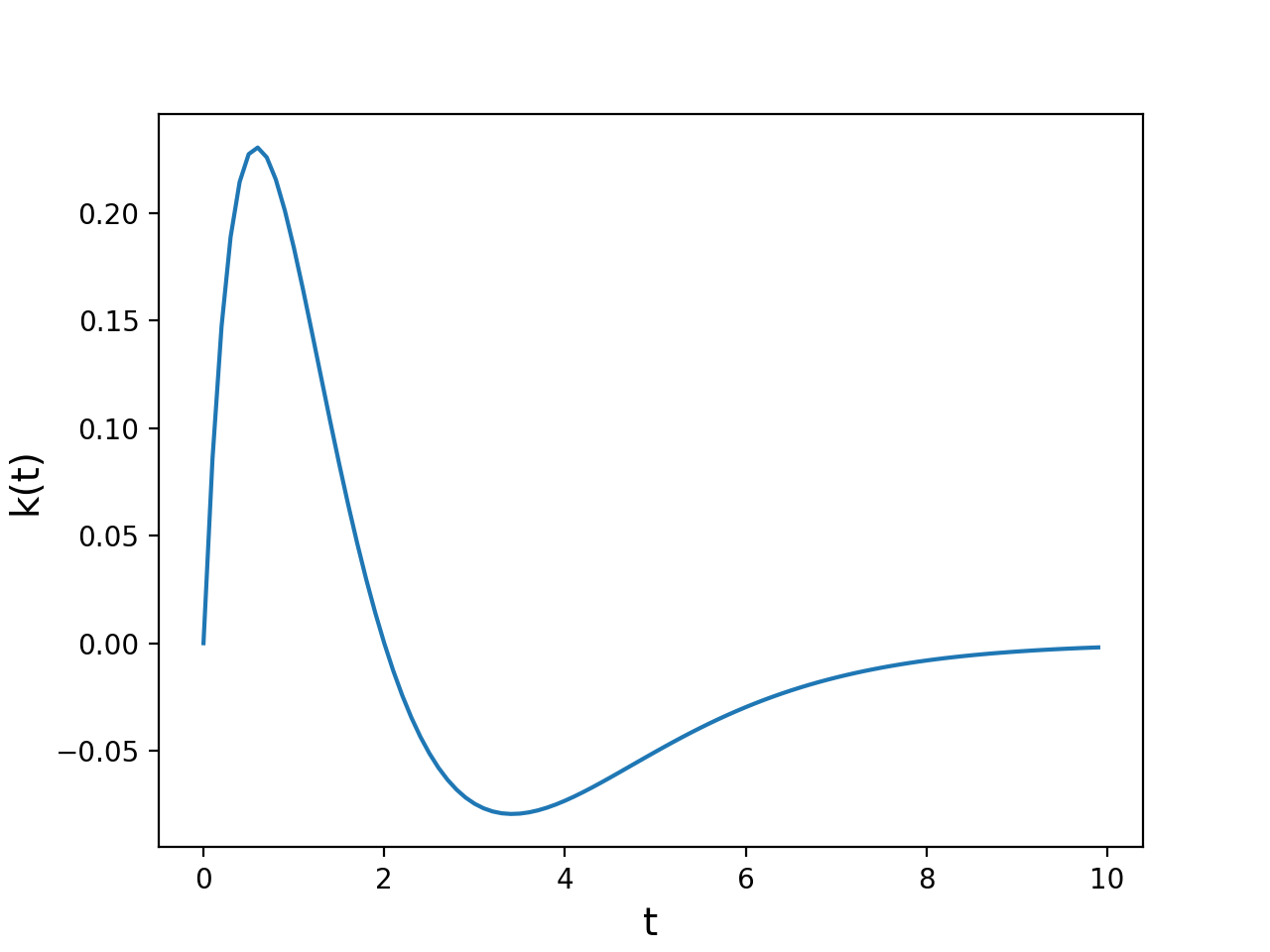}%
\label{fig:contkt}}
\subfloat[Discrete samples of $k\lbrack n \rbrack$ for $n$ from 0 to 4 and for $a=0.5$]{\includegraphics[width=2.5in]{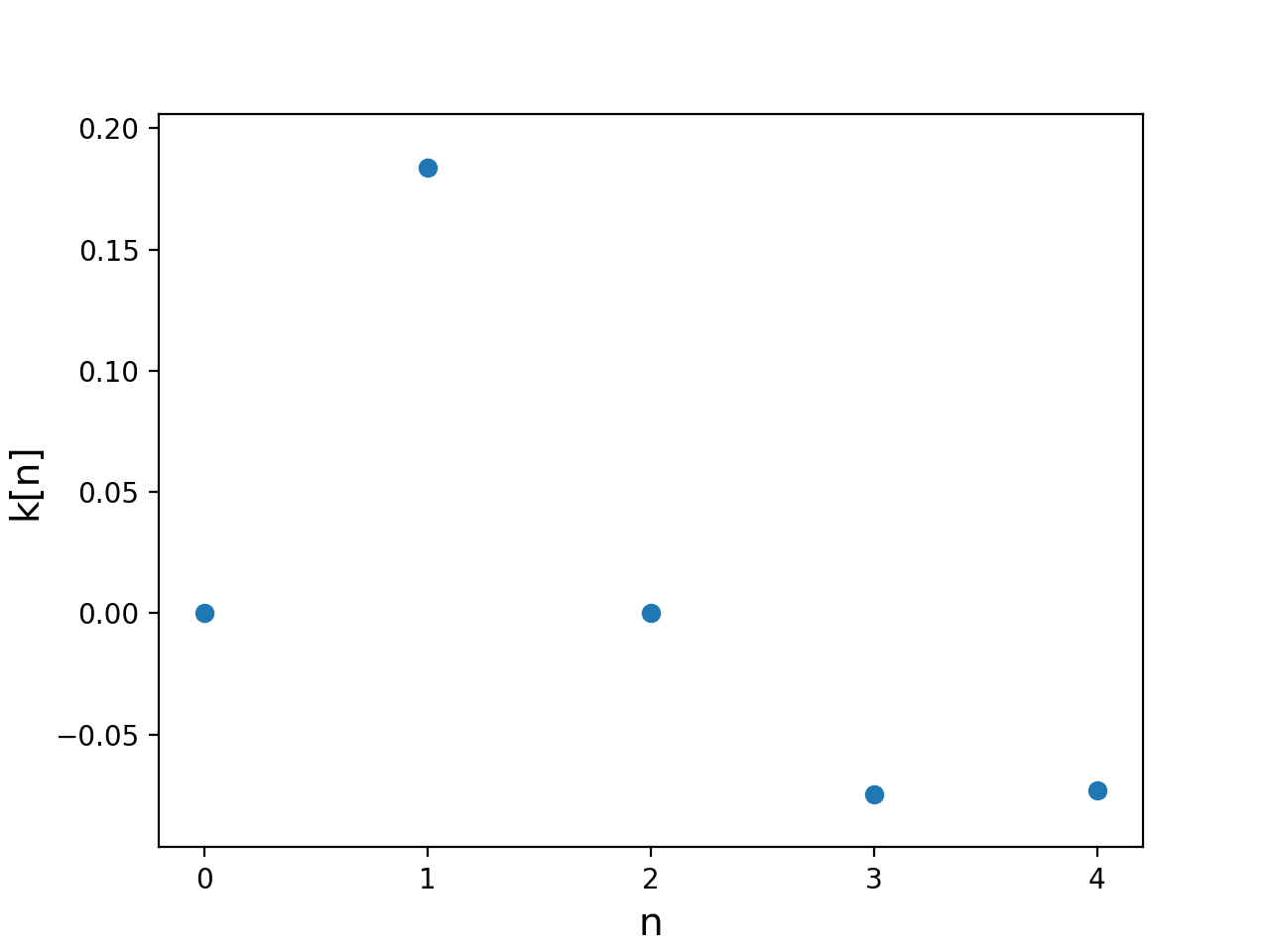}%
\label{fig:discretefilt}}
\caption{Continuous and discrete versions of the temporal filter. We use 5 discrete coefficients for the filter in our implementation.}
\label{fig:kt}
\end{figure*}

TLVQM \cite{tlvqm} is a recent NR VQA algorithm that defines a number of distortion-specific and motion-related features that are relevant to video quality. It does this in two stages. In the first stage, a number of low-complexity features are computed on every frame. These features capture the intensity and spread of motion vectors and also include specific features responsive to blockiness, blur, and interlacing.  In the second stage, high-complexity features are computed on one frame each second. These include features that are tailored to capture underexposure, overexposure, noise, blur, blockiness, low contrast, interlacing, low sharpness, low brightness, and low colorfulness.  TLVQM has many distortion-specific features and can hence be used as a general purpose NR VQA algorithm in many settings, but it also has many parameters that must be tuned. It represents a different paradigm from natural video statistics-based models, since it does not attempt to model naturalness but instead explicitly models specific distortions. CNN-TLVQM~\cite{cnntlvqm} is a variant of this method where deep features are added to TLVQM features to obtain better performance.

MMSP-VQA \cite{mmspvqa} is a deep learning based approach for NR VQA. It was trained on a very large-scale dataset called FlickrVid-150k. Features are extracted on each frame from multiple layers of an Inception-Resnet-v2 \cite{resnet} network pretrained on ImageNet \cite{imagenet}. The features were then averaged and trained with a deep neural network. FlickrVid-150k and the source code for MMSP-VQA have not yet been released. 

NR Image Quality Assessment (IQA) algorithms have been found to be quite competitive with NR VQA algorithms on user-generated content (UGC). This is because UGC is dominated by spatial distortions and does not usually present much temporal variation on quality. NR IQA algorithms such as FRIQUEE \cite{friquee} and HIGRADE \cite{higrade} have been found to outperform NR VQA algorithms on datasets such as LIVE VQC \cite{vqc}, Konvid-1k \cite{konvid}, and YouTube-UGC \cite{konvid}. FRIQUEE uses a bag of perceptually motivated statistical features from different spaces, including luminance, color, and gradients. HIGRADE models the statistics of the log-derivative of gradients, and was designed for HDR content, but has been found to work well on UGC content as well. BRISQUE \cite{brisque} is an earlier NR IQA algorithm that utilizes a spatial model of the statistics of distorted pictures. BRISQUE models the statistics of bandpass, divisively normalized coefficients of images, based on the observation that bandpass, divisively normalized pristine images follow a first order Gaussian distribution. Distorted images change these statistics, and the parameters of fits to the statistics of an image can be used to reliably predict the quality of an image. The statistical models of distorted pictures discovered in BRISQUE underpin subsequent advances in NSS-based IQA research. NIQE \cite{niqe} also models spatially bandpassed coefficients, but does not require training. NIQE quantifies the deviation of the statistics of an image via a statistical fit to a small corpus of high-quality natural images. CORNIA \cite{cornia} is an NR IQA algorithm that does not attempt to model the statistics of natural images, but instead uses a dictionary to effectively represent images for quality assessment. Li et al.~\cite{dingquan} proposed a CNN based method for UGC quality assessment and trained it on a combination of three major UGC databases.

Space-time slices are cuts of a video through space and time along fixed, pre-determined directions and spanning an entire video. Space-time slices have been effectively used for FR VQA, but are only applicable to stored videos that are available to an algorithm in their entirety~\cite{frsts,sts2,sts3,sts1}  Space-time chips significantly modify this concept, since they are highly localized in space and time, are sensitive to local motion, and can be used in real-time applications. Space-time chips were first introduced in ChipQA-0~\cite{chipqa0}, but were found using optical flow in that method. Optical flow is generally expensive to compute, and their requirement can make algorithms computationally impractical, although motion is relevant to any study of video quality. Motion has also been used in several FR VQA models~\cite{movie,motionstruct,frof} and NR VQA models~\cite{vbliinds,chipqa0,tlvqm,nrof}. In our work, we use implicit motion to define the ST Chips used in quality prediction, based on a simple regularity maximizing concept and without using optical flow. {Our new model is able to obtain better performance with much lower computational complexity than the original prototype ChipQA-0.} ChipQA-0 also performed poorly on UGC databases, while the full model, ChipQA, incorporates temporal filtering, and utilizes chroma and gradient features, yielding a holistic algorithm that performs well on both professional and  user generated content.

\section{Video quality assessment using space-time chips}

\subsection{Space-Time Perception}
When a video signal is incident on the retina, it is subjected to bandpass spatial filtering expressed at the outputs of the retinal ganglion cells. In a simple model of this process, local spatial averages of the signal are subtracted from the signal, and a form of adaptive gain control is applied on the difference~\cite{simoncelli}. The resultant signal has a greatly reduced entropy and is carried by the optical nerve at a reduced bandwidth to further stages along the visual pathway. This ``contrast signal" is subsequently subjected to temporal entropy reduction filtering (~\cite{simoncelliMT,lgn, adelson, ahumada}) which can also be modelled in a simple way as a temporal bandpass filter operation, with filter kernel given by
\begin{equation}\label{eq:kt} 
k(t) = t(1-at)\exp (-2at)u(t),
\end{equation}
where $t$ denotes time, $a$ is a constant parameter, and $u(t)$ is the unit step function. The function is plotted against $t$ in Fig.~\ref{fig:contkt}.
These processes serve to spatially and temporally decorrelate the visual signal. These initial stages of the human visual system motivate the use of spatial and temporal decorrelating functions on videos before analyzing their statistics. 

When the visual signal arrives at area V1 (the primary visual cortex) it is decomposed into orientation and scale-tuned spatial and temporal channels. Neurons in area V1 are also sensitive to specific local orientations of motion. From here, the visual signal is passed to area middle temporal (MT) in extrastriate cortex, where further motion processing occurs~\cite{simoncelliMT,movie}. MT contains neurons sensitive to motion over larger spatial fields, and the neural representation of the space-time visual signal at this point makes efficient use of space-time regularity. Similarly, ST Chips are sensitive to local orientations of motions aggregated over large spatial fields, which we use to build spatiotemporal representations of video data. 

\subsection{Defining Space-Time Chips}
The first step in our algorithm is to compute spatial mean-subtracted and contrast-normalized (MSCN) coefficients of each frame in a given video. Given a luminance image $I\lbrack i,j,n\rbrack$ at frame index (time) $n$, the MSCN coefficients $\hat{I}\lbrack i,j,n\rbrack$ are:
\begin{equation}\label{eq:mscn} 
\hat{I}\lbrack i,j,n \rbrack = \frac{I\lbrack i,j,n\rbrack-\mu\lbrack i,j,n\rbrack}{\sigma\lbrack i,j,n \rbrack +C}    
\end{equation}
where $i\in 1,2..M$, $j=1,2..N$ are the spatial indices, $M$ and $N$ are the height and width of the image respectively, $C$ is a constant for numerical stability, and
  \begin{equation}\label{eq:mean} 
  \mu\lbrack i,j,n\rbrack = \sum\limits_{k=-K}^{k=K} \sum\limits_{k=-K}^{k=K} w\lbrack k,l\rbrack I\lbrack i+k,j+l,n\rbrack
  \end{equation}
\begin{equation}\label{eq:sigma} 
\sigma\lbrack i,j,n\rbrack = \sqrt{\sum\limits_{k=-K}^{k=K} \sum\limits_{k=-K}^{k=K} w\lbrack k,l\rbrack (I\lbrack i+k,j+l,n\rbrack-\mu\lbrack i,j,n\rbrack)^2}
\end{equation}
are the local spatial mean and standard deviation of luminance, respectively. $w = \{w\lbrack k,l\rbrack, k\in -K,..,K, l\in -K,..,K\}$ is a 2D circularly-symmetric Gaussian weighting function sampled out to 3 standard deviations and rescaled to unit volume. We use $K=3$ in our implementation. Research on natural image statistics \cite{brisque, friquee,higrade} has shown that, in the absence of distortion, the coefficients $\hat{I}\lbrack i,j,n \rbrack$ can be expected to reliably follow a first-order generalized Gaussian distribution (GGD). This is the basis of state-of-the-art NR IQA algorithms such as FRIQUEE, HIGRADE, and BRISQUE, and is used in VQA algorithms as well to model spatial statistics, e.g., in V-BLIINDS. Moreover, the MSCN operation supplies a reasonable approximation to bandpass processing and adaptive gain control (relevant to contrast masking) that occurs in the retina. 

Following spatial MSCN processing, we apply the causal temporal filter (Eq.\ref{eq:kt}) to groups of $T'$ consecutive frames, with no overlap between adjacent groups of frames. This is cheaper than using overlapping blocks and we found that this does not affect performance. The discrete coefficients of the filter, $k[n]$, are shown in Fig.~\ref{fig:discretefilt} for $a=0.5$ and the length of the filter $P=5$. We experimented with different values of $a$ and discuss how they affect performance in the results section. We denote the result of this temporal operation as $D$.
\begin{equation}
D\lbrack i,j,n\rbrack = \hat{I}\lbrack i,j,n\rbrack*k\lbrack n\rbrack
\end{equation}
We use reflective padding in the temporal dimension at the boundaries of each block of $T'$ frames such that the output also has $T'$ frames. We fix $P=T'$ to minimize the effect of boundary artifacts, as increasing P to be greater than T would result in a greater use of padded points. For ease of representation, denote the processed frame $D\lbrack i,j,n\rbrack$ for $i\in 1,2..M$, $j=1,2..N$ at time instance $n$ as $D_n$. 

\begin{figure} 
  \includegraphics[width=\linewidth]{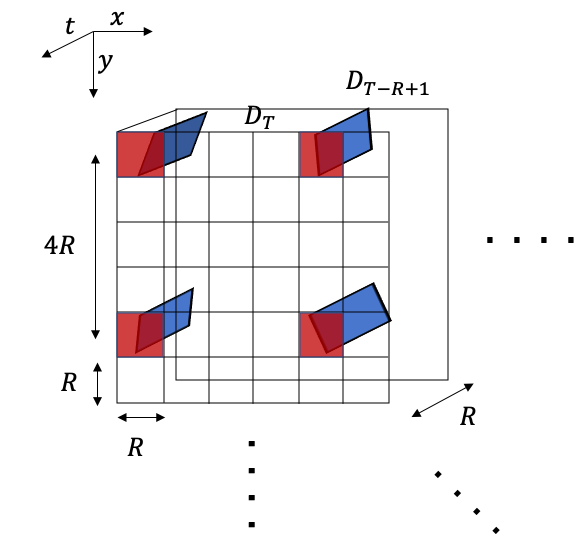}
  \caption{Extracting ST Chips from a video volume of spatially and temporally decorrelated frames from $D_{T-R+1}$ to $D_T$. One portion of the video is shown for illustration. ST Chips are extracted by cutting through the volume over $R\times R$ windows (that are spaced apart by $4R$ pixels) $R$ frames back in time. ST Chips are the angled squares in blue, and the windows are shown in red. }\label{fig:extract_stchips}
\vspace{-5mm}
\end{figure}

\begin{figure} 
  \includegraphics[width=\linewidth]{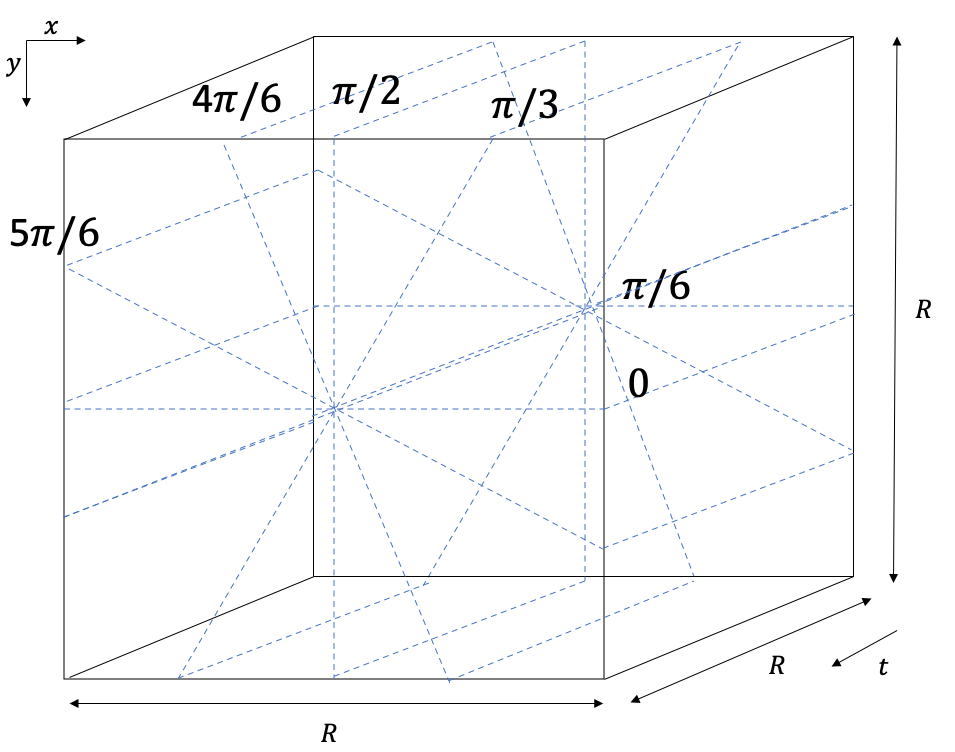}
  \caption{Finding the best ST-Chip over a particular $R\times R$ spatial window. Chips are extracted from a $R\times R\times R$ video volume along 6 angles that are equally spaced from 0 to $\pi$. The angles are shown next to their corresponding chips. The chip which has the minimum excess kurtosis is selected as the chip that best captures motion. This criterion is based on the Gaussianity of natural images and the smoothness of motion.}\label{fig:stdir}
  \vspace{-4mm}
\end{figure}

\begin{figure*}
\centering
\subfloat[Aliased and pristine]{{\includegraphics[width=0.24\textwidth]{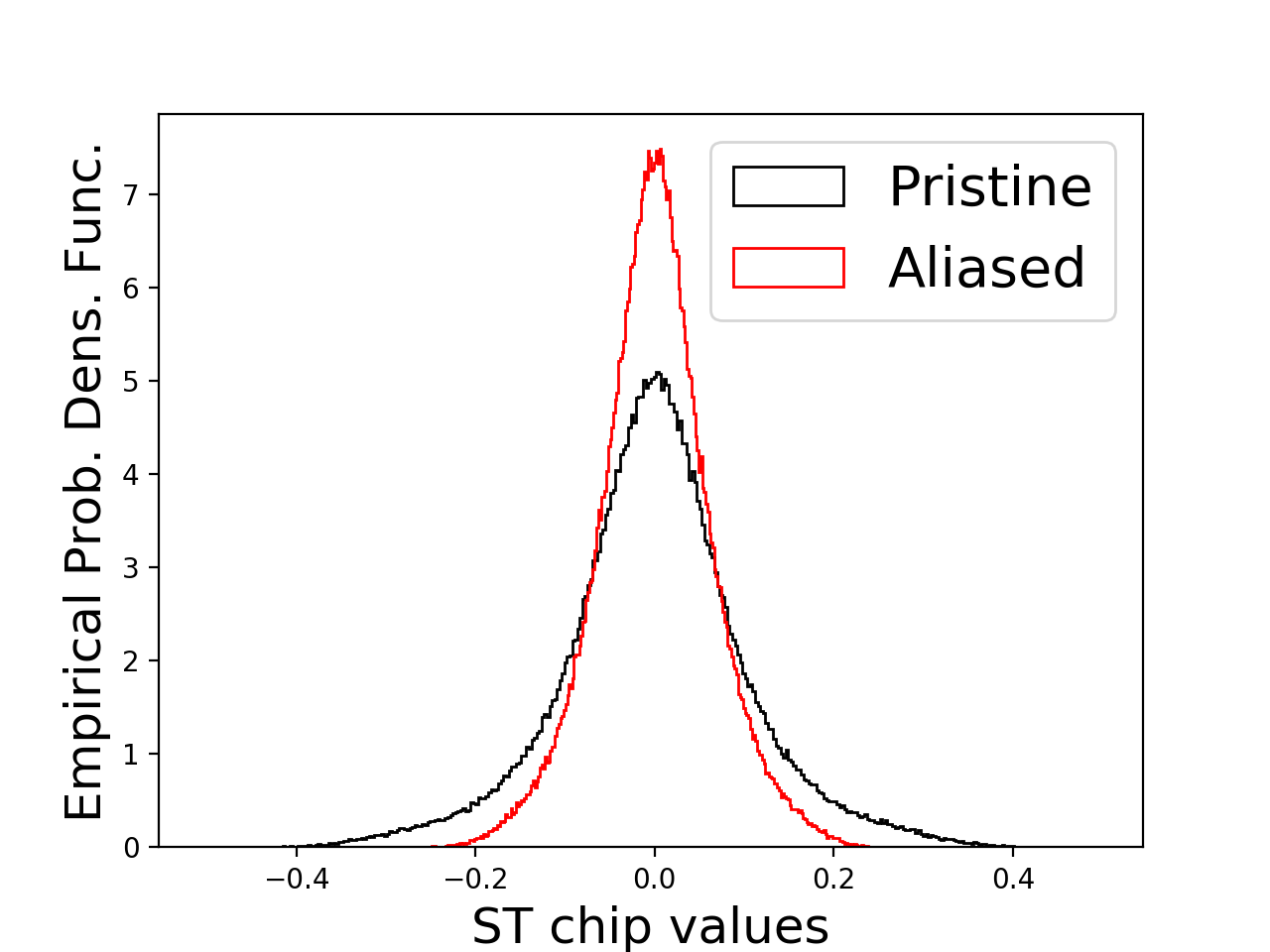}}}
\subfloat[Compressed and pristine]{{\includegraphics[width=0.24\textwidth]{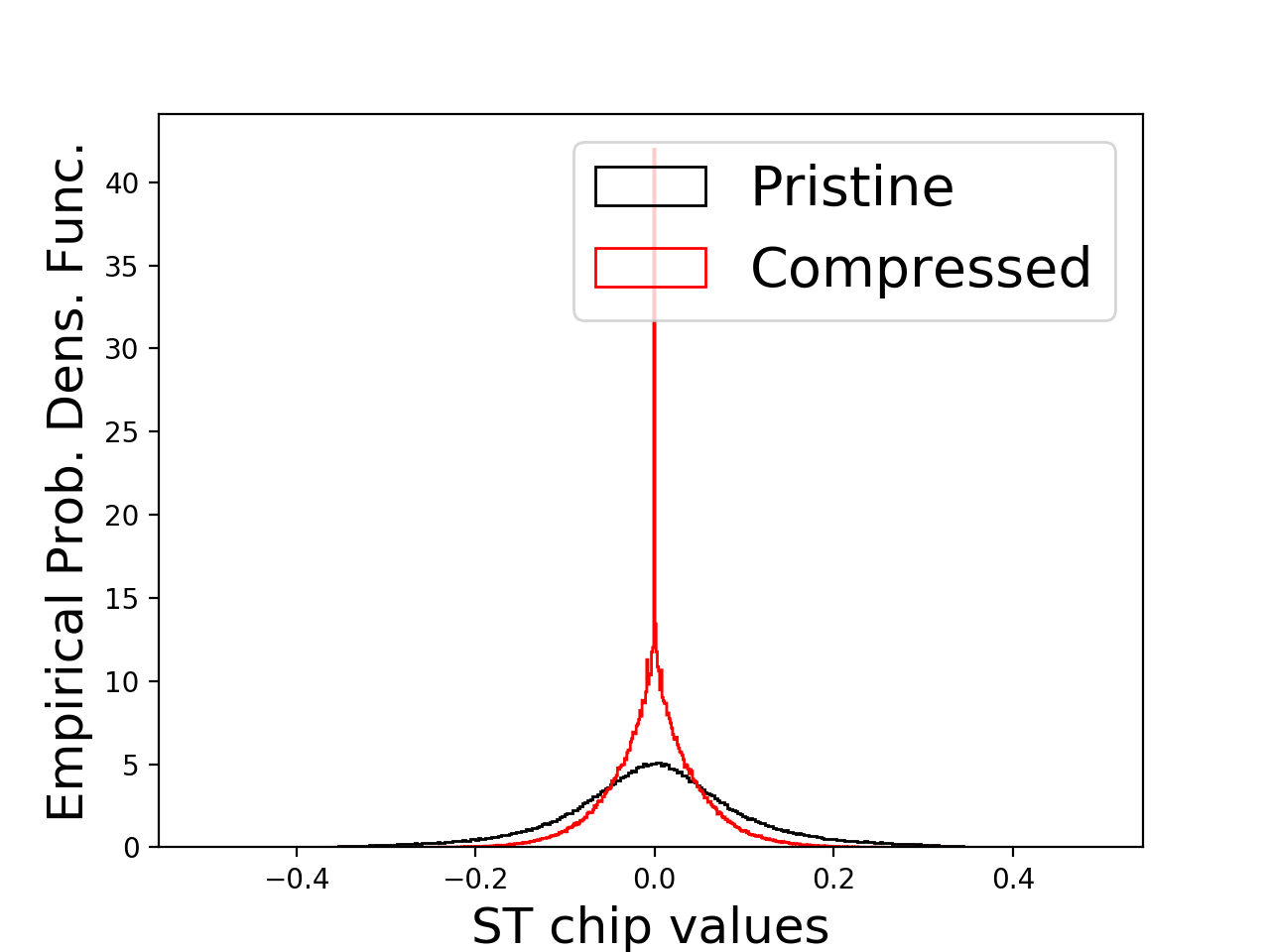}}} 
\subfloat[Flicker and pristine]{{\includegraphics[width=0.24\textwidth]{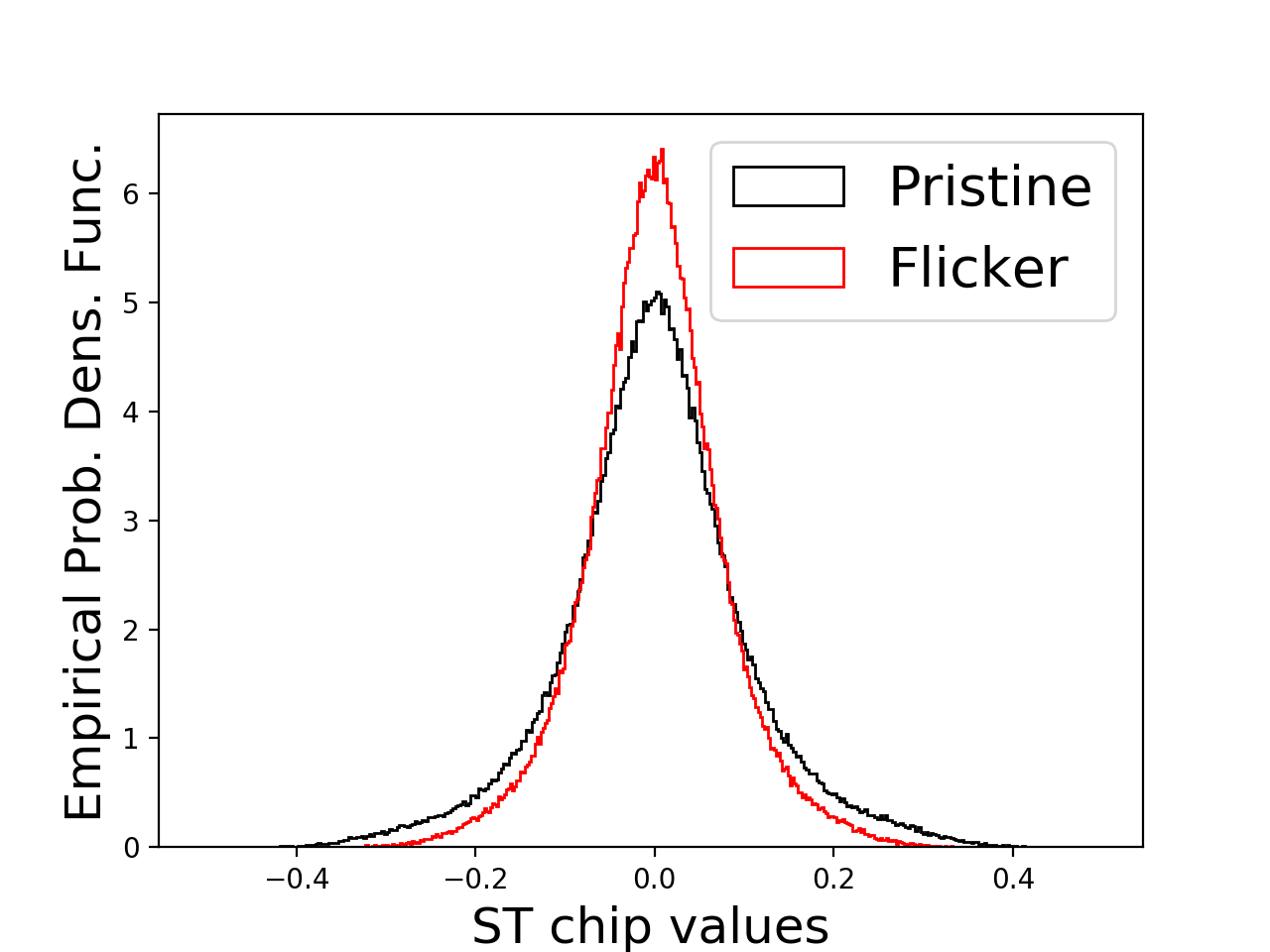}}}
\subfloat[Interlacing and pristine]{{\includegraphics[width=0.24\textwidth]{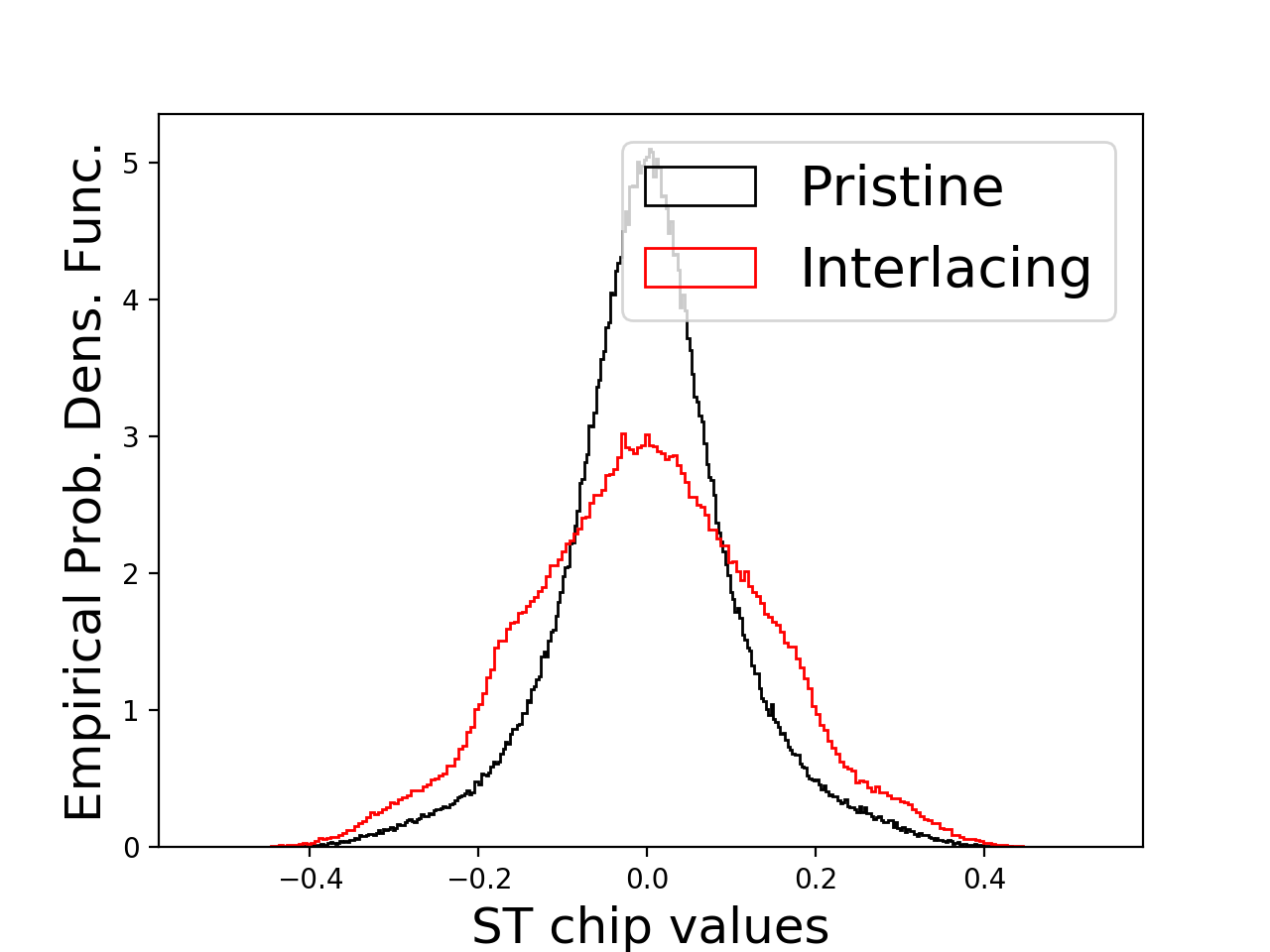}}}
\caption{Empirical distributions of ST Chips. Pristine (original) distributions are in black and distorted distributions are in red.  }
\label{fig:stchips}
\centering
\subfloat[Flicker and pristine]{{\includegraphics[width=0.24\textwidth]{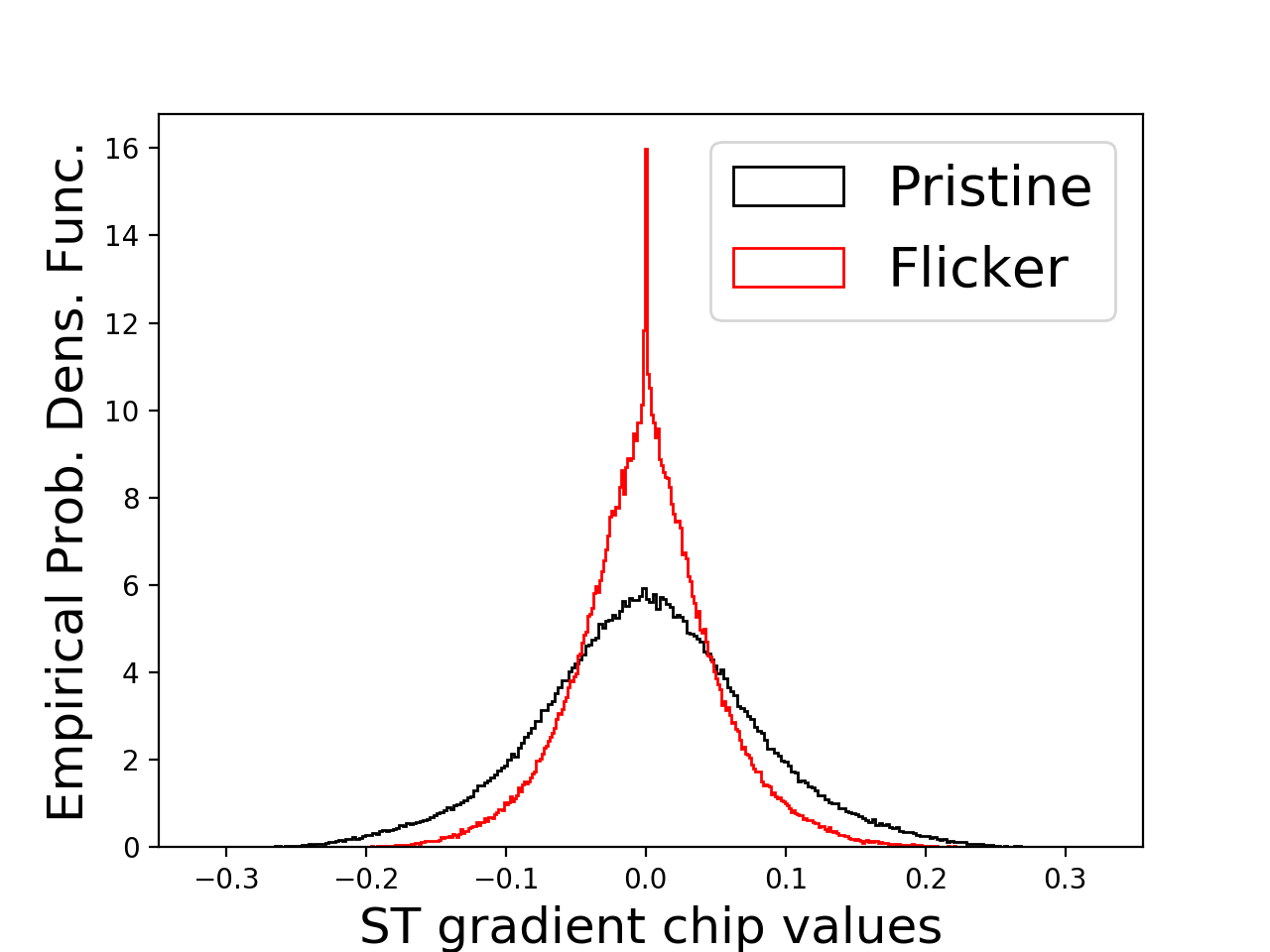}}}
\subfloat[Frame Drop and pristine]{{\includegraphics[width=0.24\textwidth]{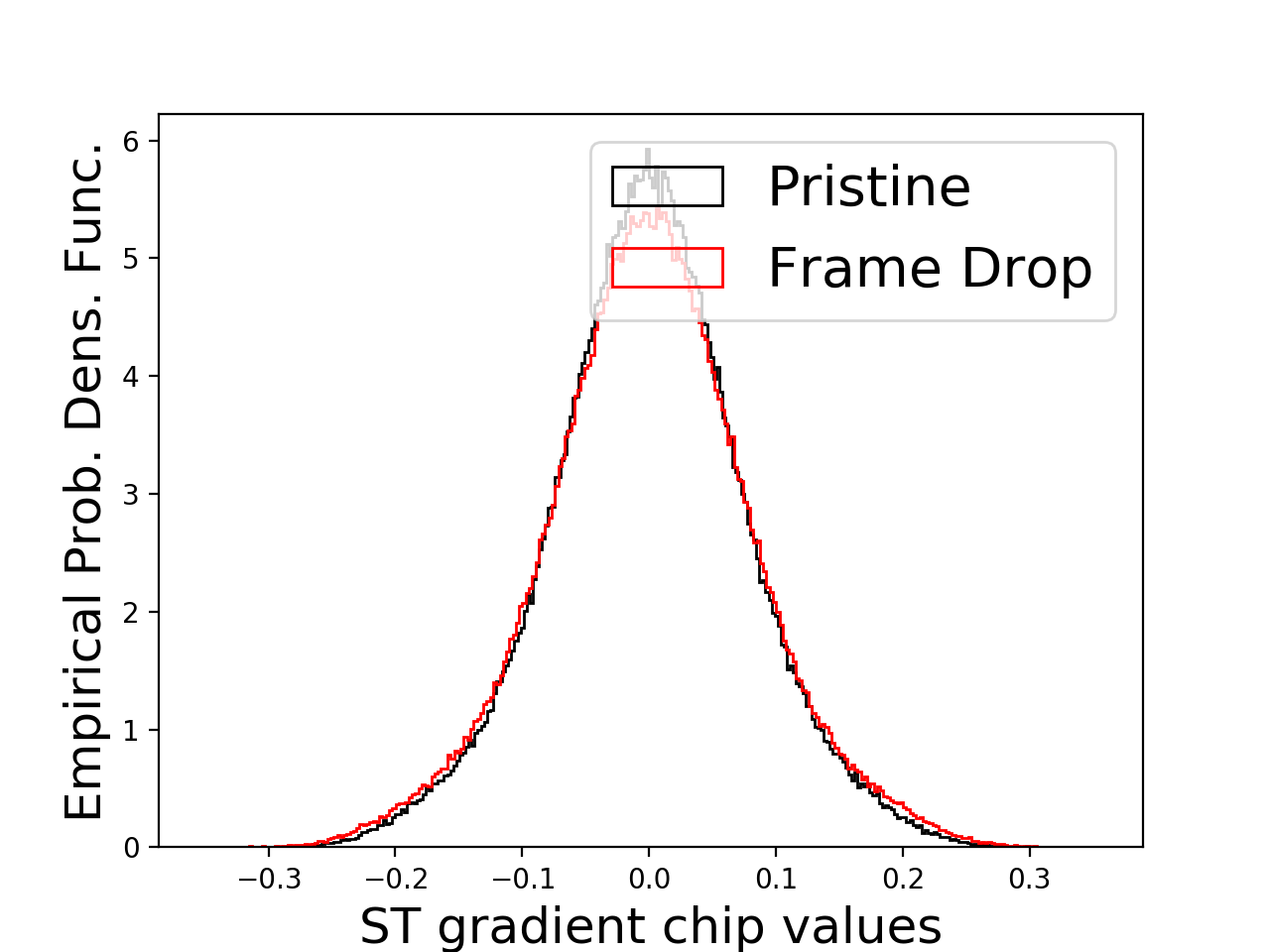}}} 
\subfloat[Interlacing and pristine]{{\includegraphics[width=0.24\textwidth]{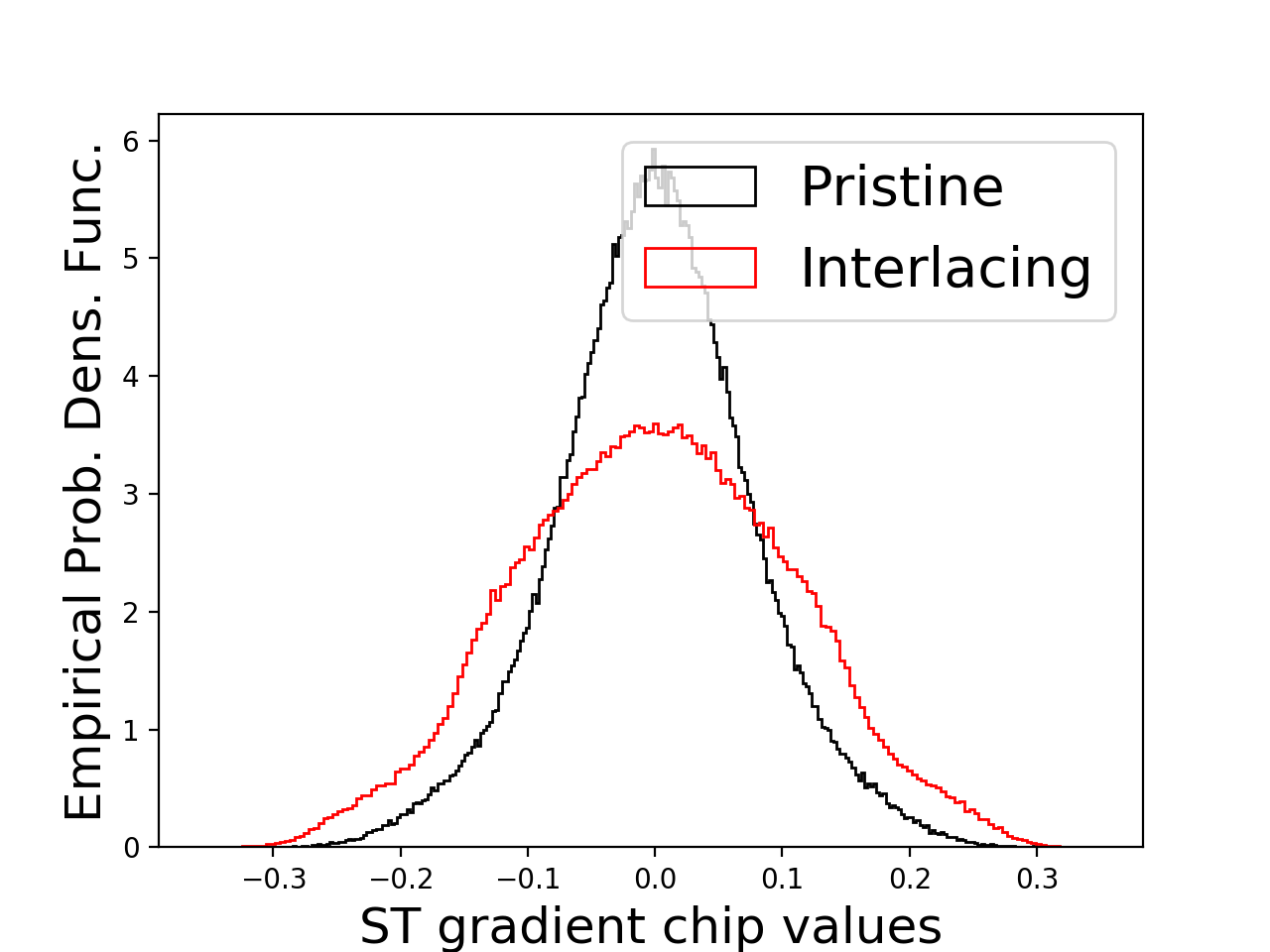}}}
\subfloat[Judder and pristine.]{{\includegraphics[width=0.24\textwidth]{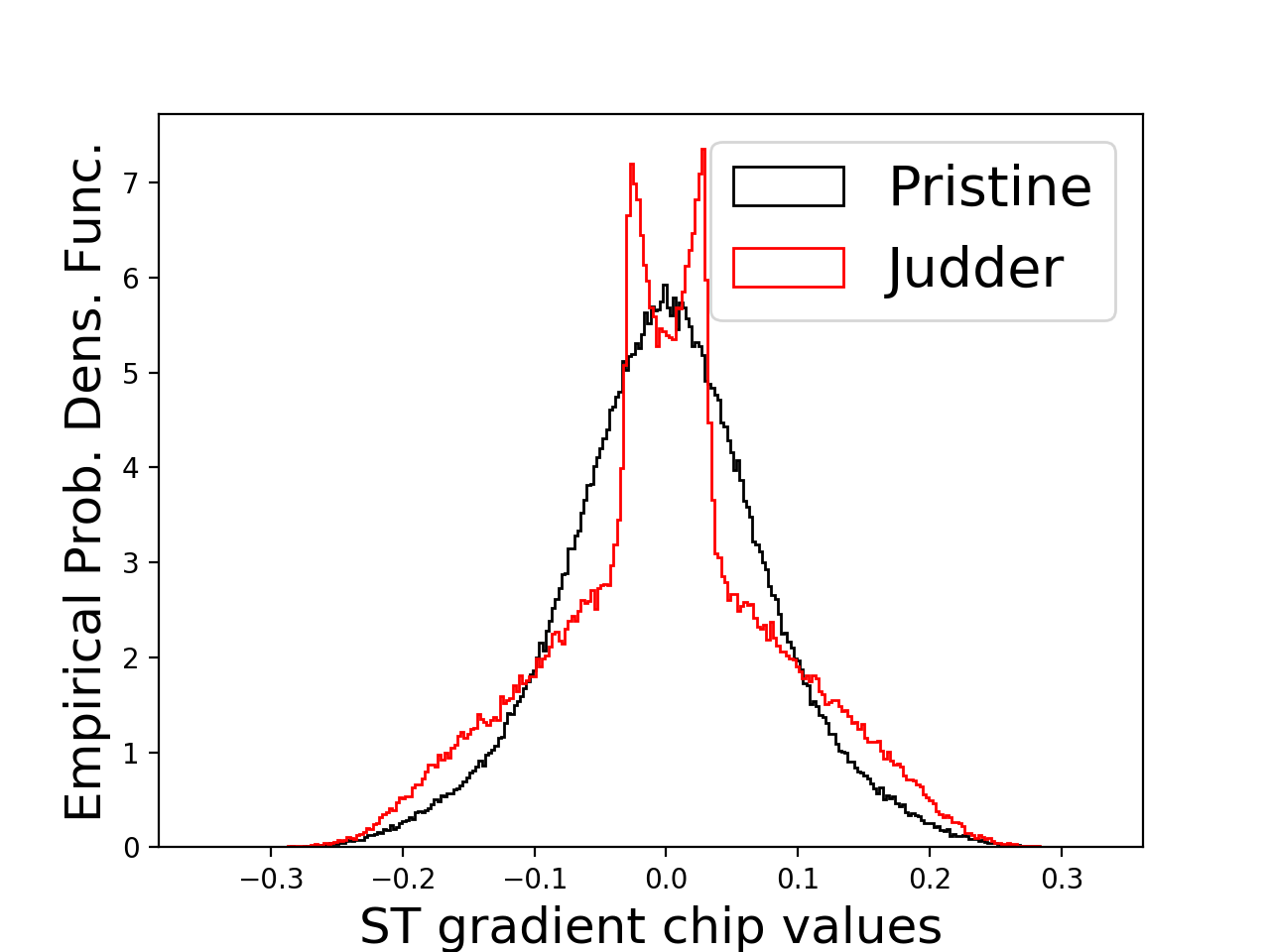}}}
\caption[]{Empirical distributions of ST Gradient Chips. Pristine (original) distributions are in black and distorted distributions are in red.}
\label{fig:stgradchips}
\centering
\subfloat[Compressed and pristine]{{\includegraphics[width=0.24\textwidth]{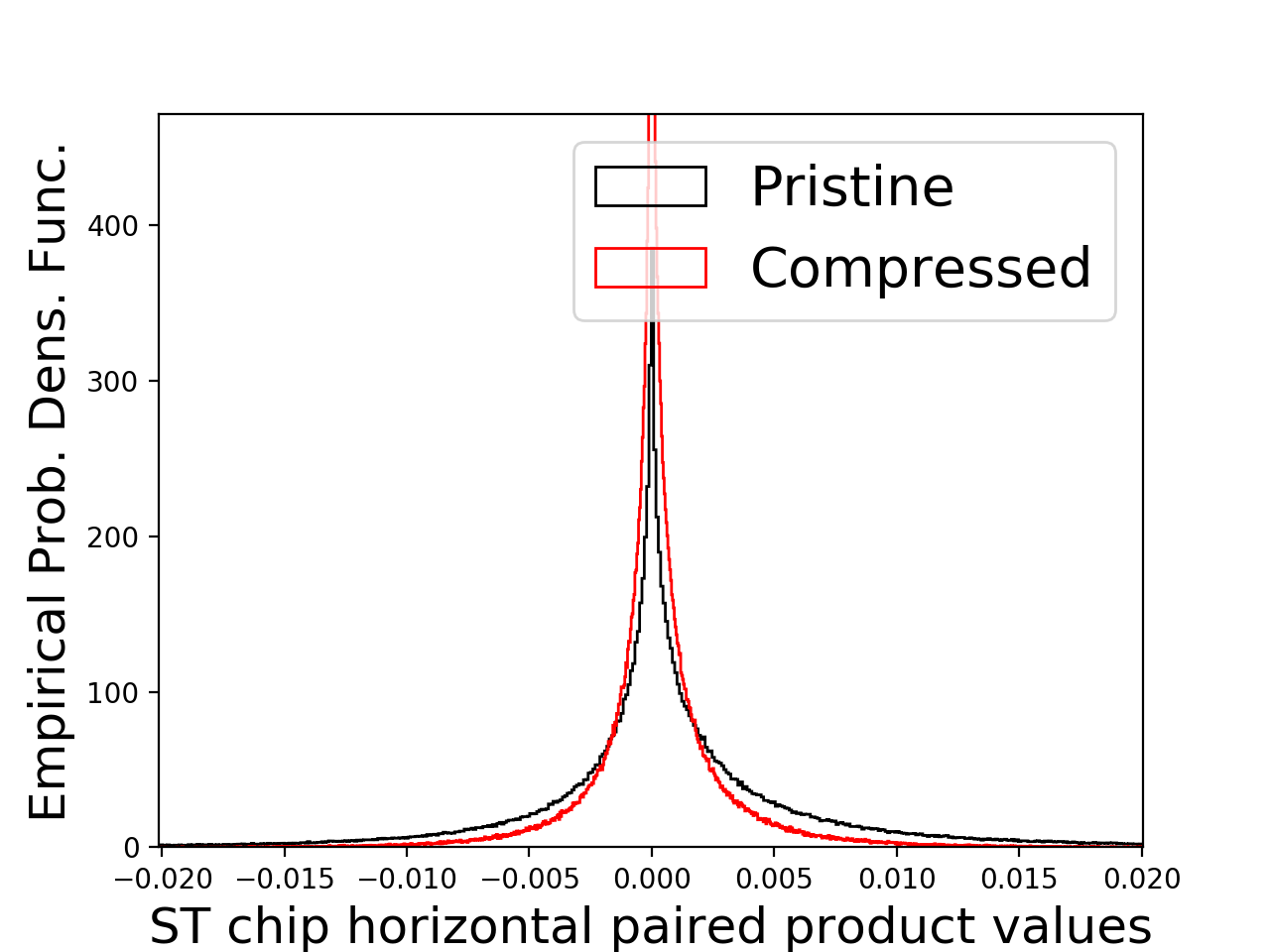}}}
\subfloat[Flicker and pristine]{{\includegraphics[width=0.24\textwidth]{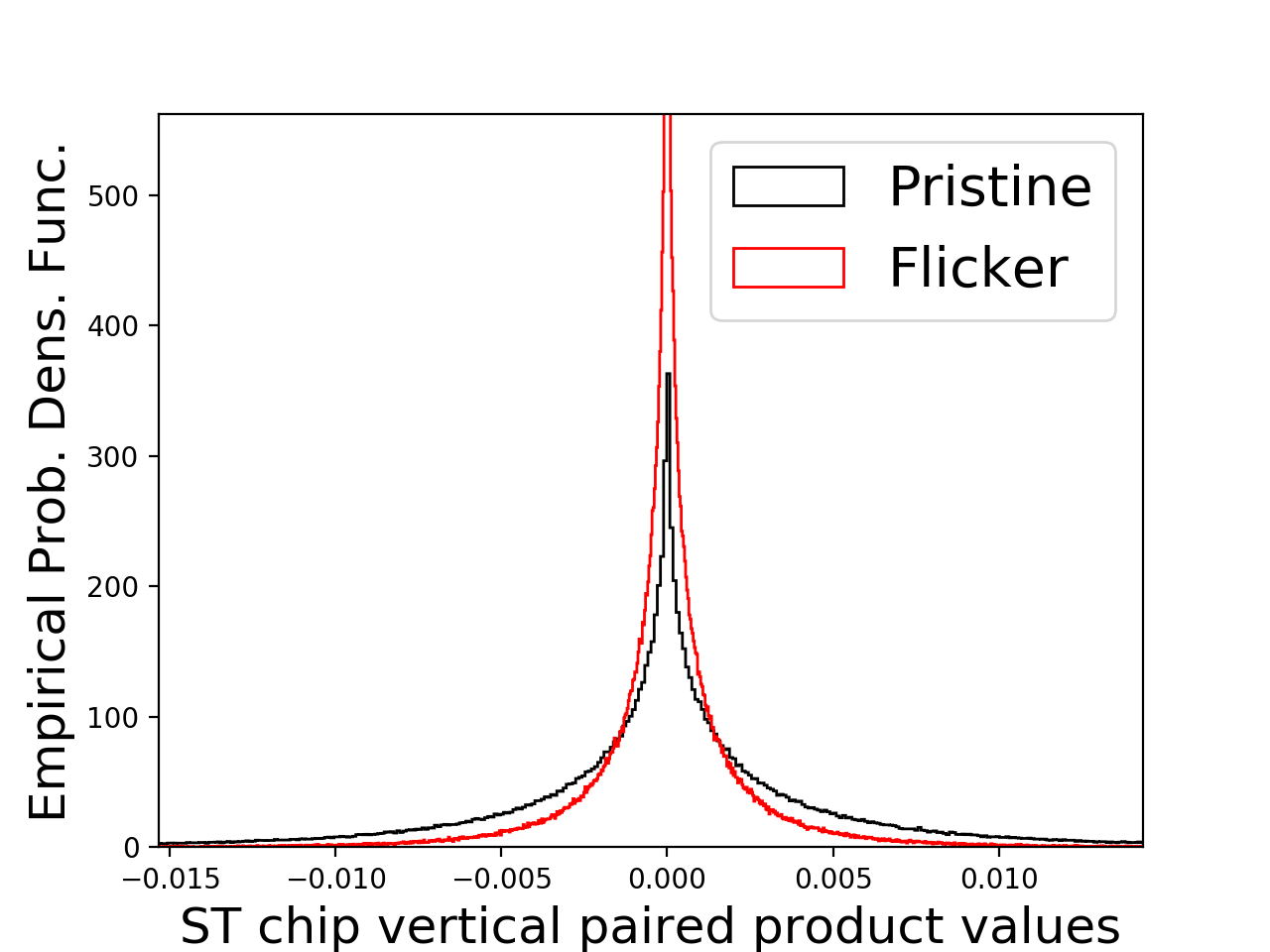}}} 
\subfloat[Interlacing and pristine]{{\includegraphics[width=0.24\textwidth]{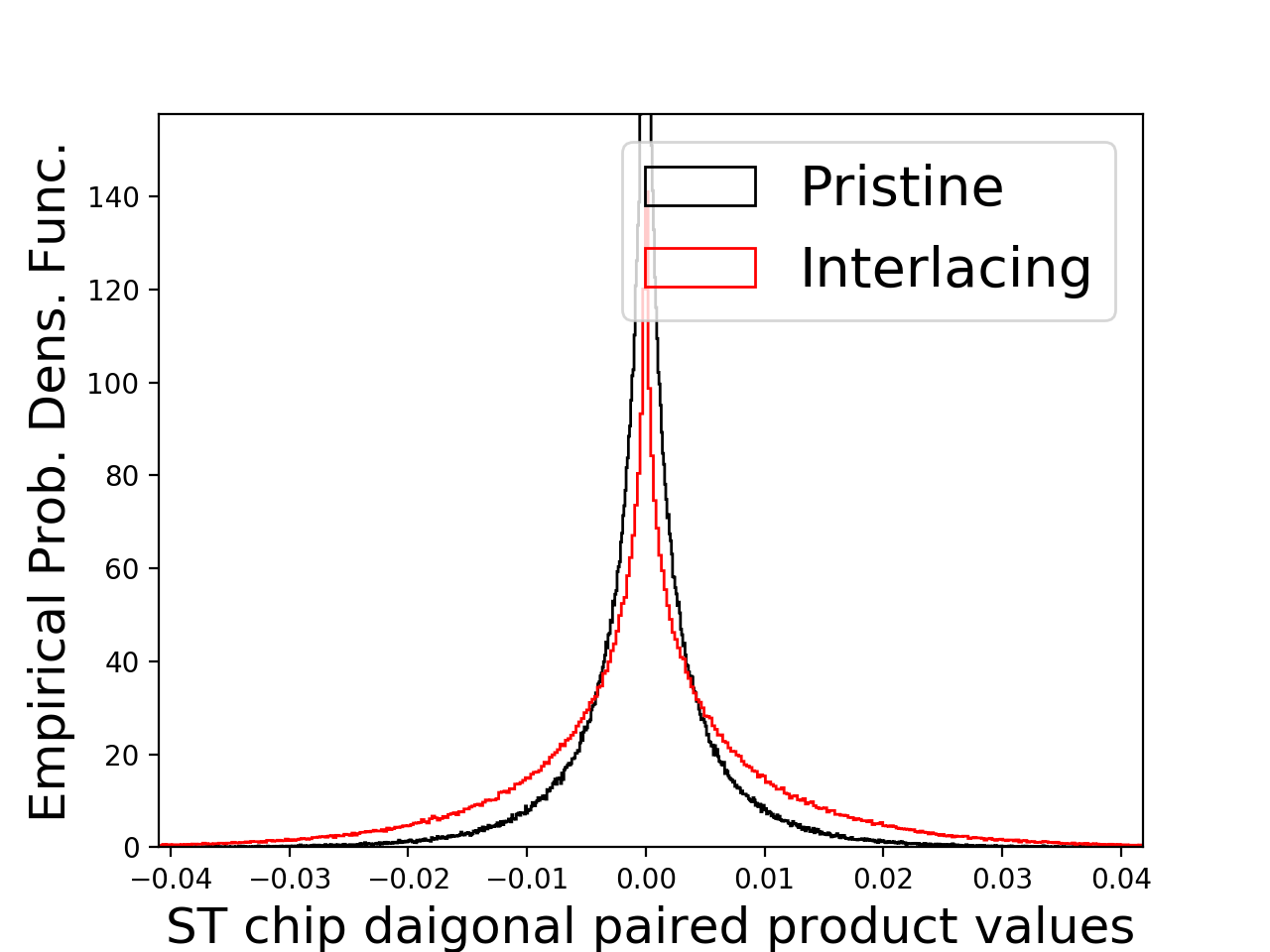}}}
\subfloat[Judder and pristine]{{\includegraphics[width=0.24\textwidth]{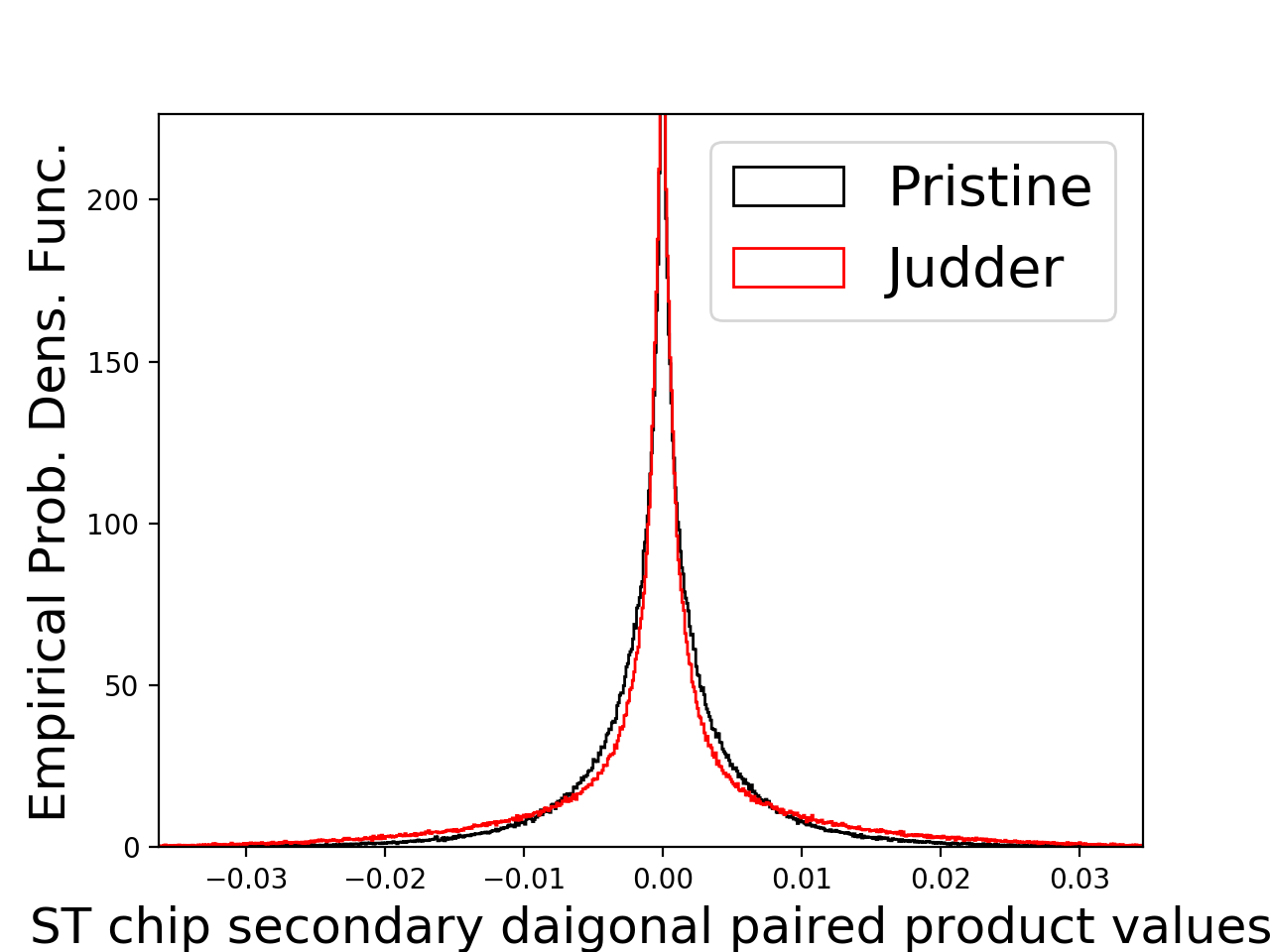}}}
\caption{Empirical distributions of paired products of ST-Chip. Pristine (original) distributions are in black and distorted distributions are in red.}
\label{fig:stpairedchips}
\centering
\subfloat[Judder and pristine]{{\includegraphics[width=0.24\textwidth]{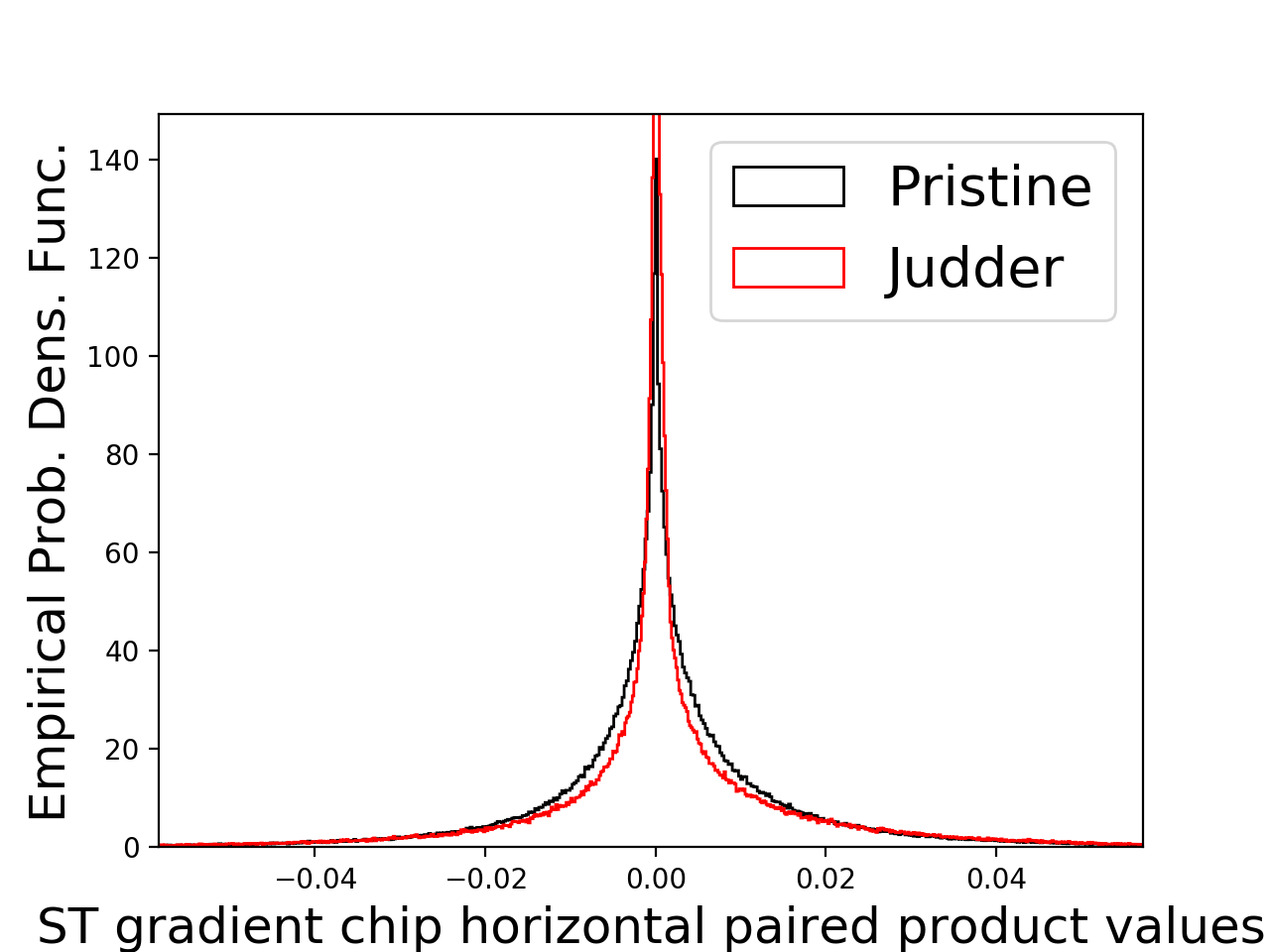}}}
\subfloat[Aliased and pristine]{{\includegraphics[width=0.24\textwidth]{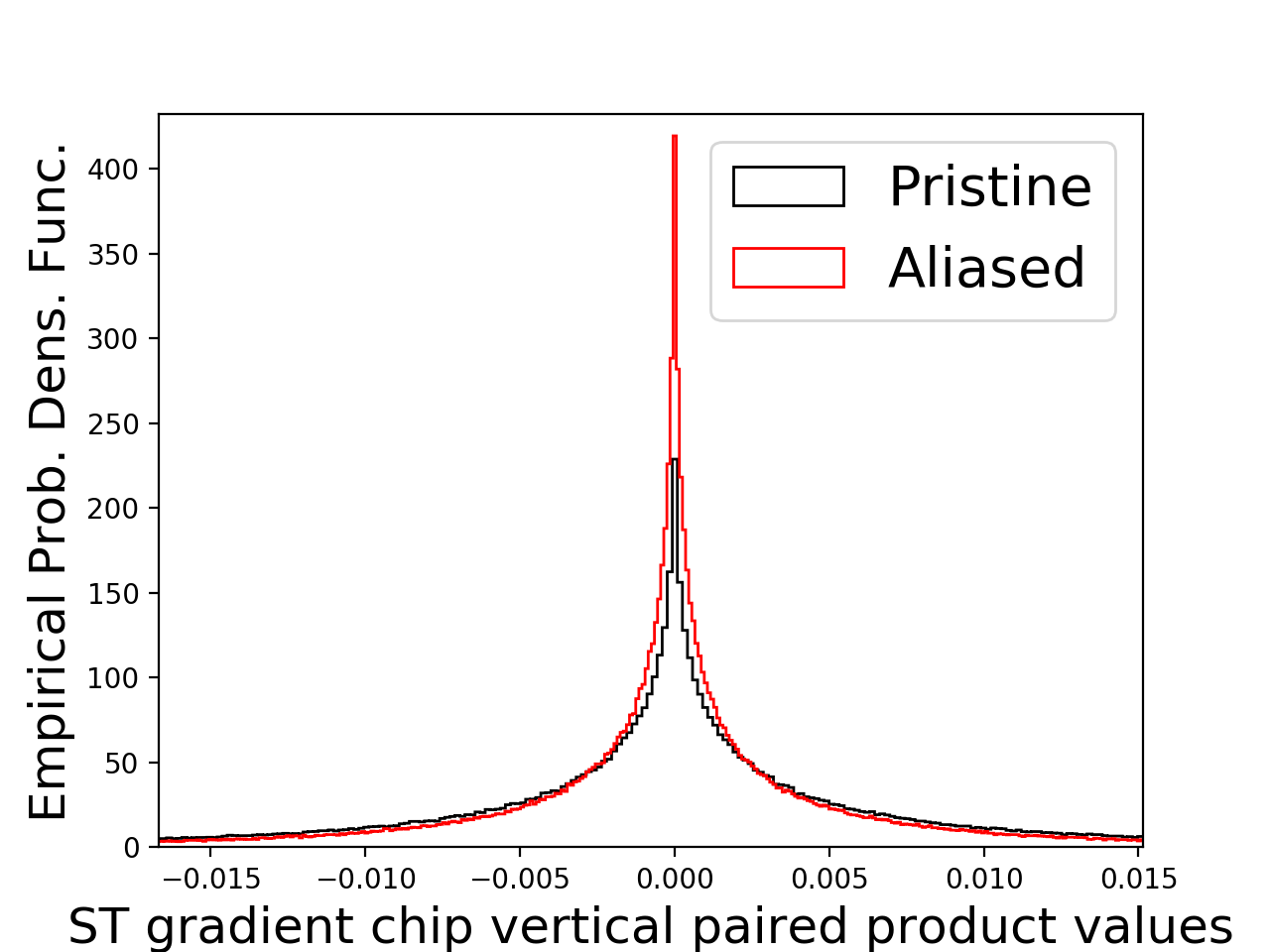}}} 
\subfloat[Compressed and pristine]{{\includegraphics[width=0.24\textwidth]{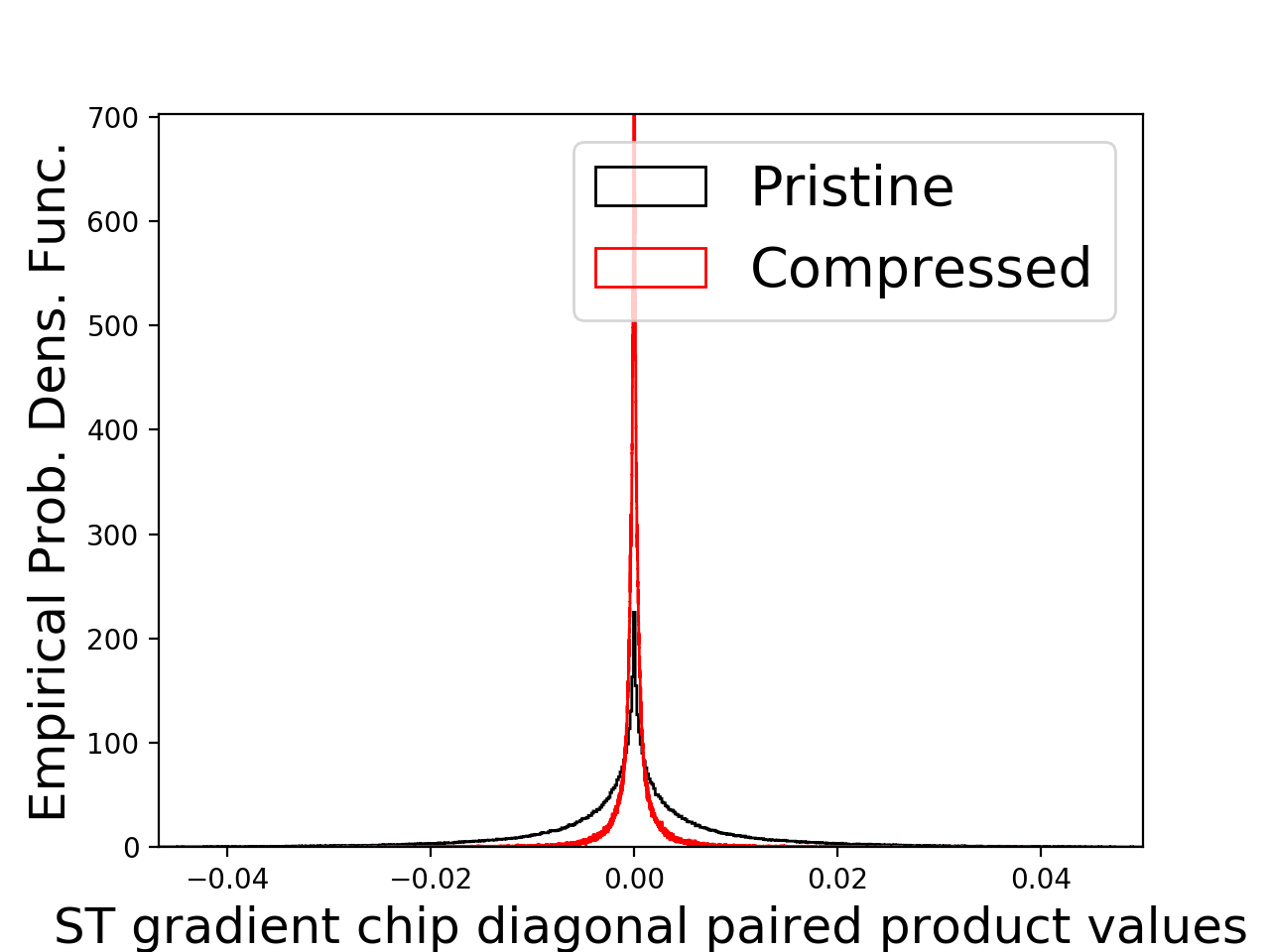}}}
\subfloat[Flicker and pristine]{{\includegraphics[width=0.24\textwidth]{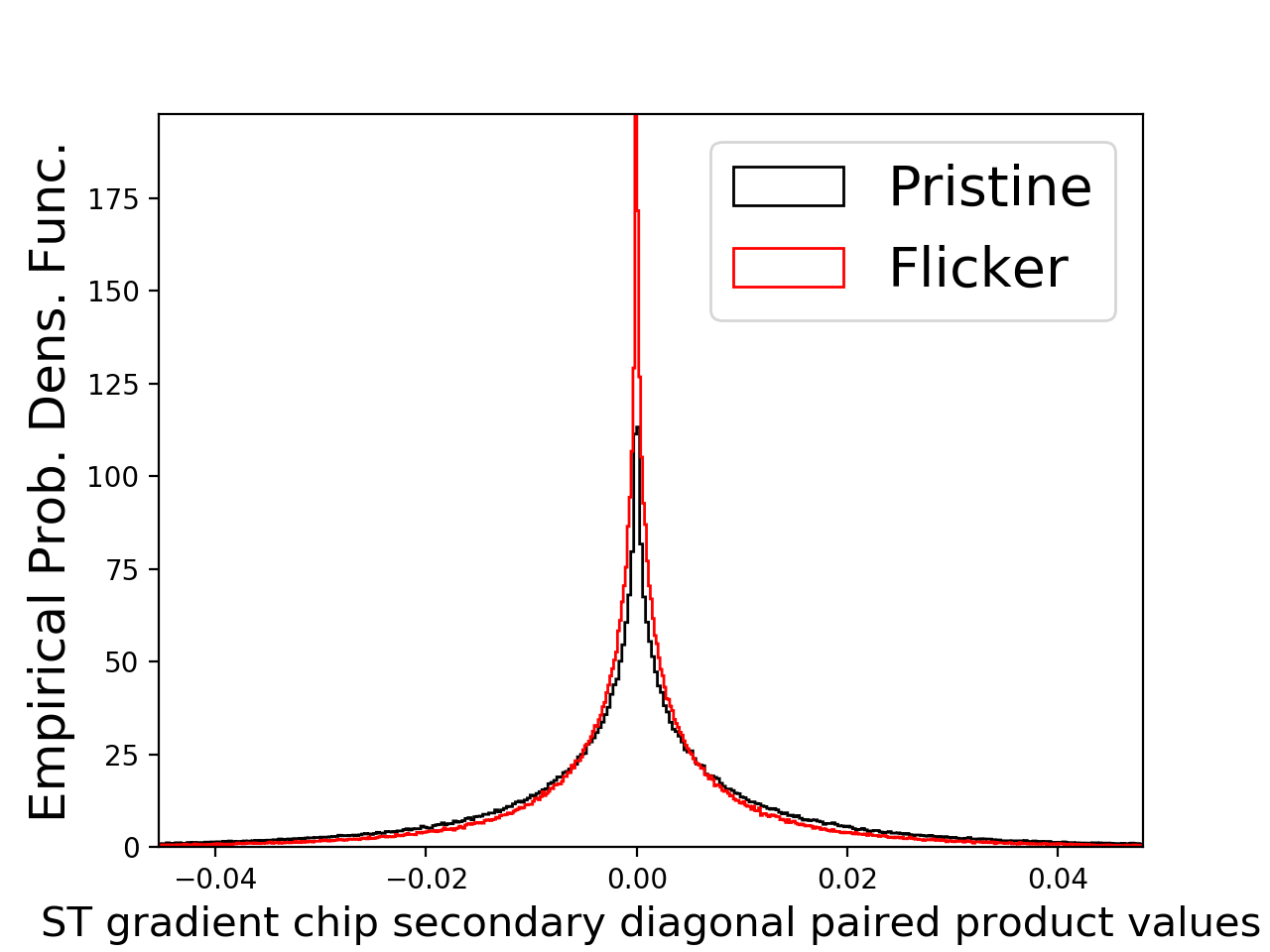}}}
\caption{Empirical distributions of paired products of ST Gradient Chips. Pristine (original) distributions are in black and distorted distributions are in red.}
\label{fig:stgradpairedchips}
\end{figure*}

We are interested in finding important directions at different spatial locations along which ST Chips can be extracted. In our experimental model, we fixed $T'=R$ for simplicity, so that each ST chip is extracted from an $R\times R\times R$ volume. At a particular time instance $T$, consider the output of the previous operation $D_n$ over indices $n=T-R+1$ to $n=T$, which is a single block of $R$ frames. Divide $D_T$ into spatial windows of size $R\times R$.  For each $R\times R$ window, we define chips that pass through the block of frames from  $D_T$  backwards in time to $D_{T-R+1}$, and that are constrained to pass through the center of the $R\times R$ window such that the normal vector to any chip lies on the $xy$ plane. Some examples of chips are shown in  Fig.~\ref{fig:stchips} (in blue, with the $R\times R$ windows in red) and chips for a single $R\times R\times R$ volume are shown in Fig.~\ref{fig:stdir}. These chips can be oriented at diverse angles. Among these angles, one is assumed to best capture the local motion, and a chip that is oriented perpendicular to the motion vector at this location will capture objects in motion along the motion vector. This is illustrated in Fig.~\ref{fig:chip_illus}. Under these constraints on the chip, we are assuming that the motion is along a vector on the $xy$ plane, which relies implicitly on the assumption that motion is linear and translational in small spatiotemporal volumes. This is a reasonable assumption that forms the basis of most modern motion estimation algorithms~\cite{horn,black}. 

In our earlier work~\cite{chipqa0}, we found the directions of motion explicitly using optical flow. This is expensive and depends on the accuracy of the optical flow algorithm. Assuming that motion is smooth for a pristine video, we expect the chips that are perpendicular to the directions of motion to follow similar statistics as natural images, since they contain projections of natural scenes as they move. The MSCNs of natural images are known to reliably obey a Gaussian law~\cite{vbliinds,brisque,niqe,simoncelli,ruderman}. Assuming the veracity of these natural image statistics models, and smoothness and linearity of motion in local regions of pristine videos, we find the directions of motions implicitly by selecting a chip having a sample set that is closest to being Gaussian amongst all of the potential chips.  This is done in a simple and direct way by computing the sample kurtosis~\cite{stats} of each chip along $Q$ different equally spaced angles, from $0$ to $\pi$, and selecting the chip that has the kurtosis closest to 3, which is the kurtosis of a Gaussian random variable, as is illustrated in Fig.~\ref{fig:stdir}. There are many possible tests of the Gaussianity of a set of chip samples. These include the Shapiro-Wilk test~\cite{shapiro}, the Anderson-Darling test~\cite{anderson}, the Martinez-Iglewicz test~\cite{martinez}, and the D'agostino kurtosis test~\cite{agostino}, which is similar to the kurtosis method that we apply. There are a number of reasons we use the simple sample kurtosis. First, we are not actually testing for Gaussianity, which these frequentist tests are designed for. Rather, we are instead ranking the chips by kurtosis and selecting which among them is most Gaussian in that sense. It is possible that all, none, or a subset of the chips may pass a given Gaussianity test, e.g., if there is little or no motion present, all may present as Gaussian. Ranking procedures on test statistics like those in~\cite{shapiro,anderson,martinez,agostino} have not been shown to measure relative Gaussianity. Further, by using the sample kurtosis we are aligning with the powerful \textit{a priori} and well-founded assumption that the bandpass chips will reliably obey a zero-mean generalized Gaussian law. The members of this distribution class may be viewed as differing \textit{only} in kurtosis, hence we may view our use of kurtosis as a conditional measure. Given the small sample size of 25, this is a powerful constraint. Lastly, the computational efficiency of the kurtosis lends it to fast implementations. We chose $Q=6$ in our implementation, and found that increasing $Q$ improves performance, although computational cost increases as well. Variation in performance as $Q$ is varied is discussed in section~\ref{results}. 

Having selected the chip that is most Gaussian at a window, we then aggregate them across windows. We do not collect chips from all $R\times R$ windows, but skip $D=4$ windows in each of the $x$ and $y$ directions. We study how performance varies with $D$ in Section~\ref{exp_impl}. The centers of the windows from which ST Chips are extracted are thus separated from each other by a distance of $4R$ pixels in each dimension. This is shown in Fig.~\ref{fig:extract_stchips}. We discuss how this spatial downsampling affects performance in Section~\ref{results} . The aggregated chips form a single ``frame'' $S$ for every group of $T'$ frames. We discretize coordinates while searching for the best chip such that each chip is of dimension $R\times R$. The dimension of the aggregated frame $S$ of ST Chips is $M'\times N'$, where $M'=\frac{R}{4}\lfloor \frac{M}{R} \rfloor$ and $N'=\frac{R}{4}\lfloor \frac{N}{R} \rfloor$. We chose $R=T'=5$ in our implementation. Variation in performance as $R$ is varied is discussed in section IV. 

We repeated the process described above for the spatial gradient magnitude field of the video as well. Gradients contain important information descriptive of edges and contrast variations and have been found to be useful for image and video quality assessment. Gradient-based features find a place in most SOTA algorithms~\cite{tlvqm,friquee,videval,higrade}.  ChipQA computes the gradient components in the vertical and horizontal directions using a Sobel kernel of size $3\times 3$. The Sobel filter eliminates low frequency information and has high-pass characteristics that detects edges. The statistics of these edges are useful for quality assessment since they are often heavily affected by distortions. We then find the MSCNs of the gradient magnitude, apply the temporal filter $k\lbrack n \rbrack$, and extract ST Chips along the directions with kurtosis closest to 3 at windows that are separated by a distance of $4R$ in each dimension. We refer to these as ``ST Gradient Chips".
\subsection{Statistics of ST Chips}
MSCNs of spatial frames and of gradient fields are known to follow regular statistics, and this is true of ST Chips of spatial frames and of gradient fields  as well because of the smoothness of motion. ST Chips and ST Gradient Chips are found to follow a generalized Gaussian distribution (GGD) of the form:
  \begin{equation}
  f(x;\alpha;\beta) = \frac{\alpha}{2\beta \Gamma(\frac{1}{\alpha})} \exp (-(\frac{|x|}{\beta})^\alpha)
  \end{equation}
where $\Gamma(.)$ is the gamma function:
\begin{equation}
\Gamma(\alpha) = \int_0^\infty t^{\alpha-1} \exp(-t) dt.
\end{equation}
The shape parameter $\alpha$ of the GGD and the variance of the distribution are estimated using the moment-matching method described in \cite{estimation}. Examples of the first-order distribution of ST Chips and ST Gradient Chips are shown in Fig.~\ref{fig:stchips} and
Fig.~\ref{fig:stgradchips} respectively. Though the ST chip is chosen to be as Gaussian as possible, previous research has shown that Gaussianity breaks in the presence of distortions~\cite{brisque,higrade}, and for ST chips the statistics could also deviate from Gaussianity for large motion fields. These deviations from Gaussianity are useful for quantifying losses in quality, and we find that the GGD is able to model these deviations in the statistics well.

We also model the second-order statistics of ST Chips.  Define the collection of ST Chips aggregated at each time instance $T$ as $S$, and define the pairwise products
\begin{align}\label{eq:pair}
\begin{split}
H[i,j,T] & = S[i,j,T]S[i,j+1,T] \\
V[i,j,T] & = S[i,j,T]S[i+1,j,T] \\
D_1[i,j,T] &= S[i,j,T]S[i+1,j+1,T] \\
D_2[i,j,T] & = S[i,j,T]S[i+1,j-1,T]
\end{split}
\end{align}
$H$, $V$, $D_1$, and $D_2$ are found to follow a asymmetric generalized Gaussian distribution (AGGD), which is given by:
\begin{equation}\label{eq:aggd}
f(x;\nu,\sigma_l^2,\sigma_r^2) = \begin{cases}
\frac{\nu}{(\beta_l+\beta_r)\Gamma (\frac{1}{\nu})} \exp(-(-\frac{x}{\beta_l})^\nu) &  x<0 \\
\frac{\nu}{(\beta_l+\beta_r)\Gamma (\frac{1}{\nu})} \exp(-(\frac{x}{\beta_r})^\nu) &  x>0 
\end{cases} 
\end{equation}
where
\begin{equation}
\beta_l = \sigma_l \sqrt{\frac{\Gamma(\frac{1}{\nu})}{\Gamma(\frac{3}{\nu})}} \quad \mathrm{and} \quad  \beta_r = \sigma_r \sqrt{\frac{\Gamma(\frac{1}{\nu})}{\Gamma(\frac{3}{\nu})}}
\end{equation}
where $\nu$ controls the shape of the distribution and $\sigma_l$ and $\sigma_r$ control the spread on each side of the mode. The parameters ($\eta,\nu,\sigma_l^2,\sigma_r^2$) are extracted from the best AGGD fit to each pairwise product, where 
\begin{equation}
\eta = (\beta_r-\beta_l) \frac{\Gamma(\frac{2}{\nu})}{\Gamma(\frac{1}{\nu})}.
\end{equation}

Empirical histograms of the paired products of ST Chips and ST Gradient Chips are shown in Fig.~\ref{fig:stpairedchips} and Fig.~\ref{fig:stgradpairedchips}, respectively. Images and videos are inherently multiscale, and distortions can manifest themselves differently at different scales. We compute ST Chips at two scales, and compute the above features at each scale. We first apply low-pass filtering to the video and then downsample it to half the original size. ST Chips are computed as described previously, and statistical features are extracted from the lower scale. 

\subsection{Spatial Features}
Spatial features are important for video quality assessment. Since user-generated content is often dominated by spatial distortions, NR IQA algorithms such as FRIQUEE and HIGRADE achieve very good performance on UGC datasets. Compression, aliasing, and interlacing are common spatial distortions that affect professional-grade content as well. In VIDEVAL~\cite{videval}, top-performing features were selected from a number of state-of-the-art algorithms. It was found that luma, color, and gradient information were vital to building a competitive VQA algorithm. Likewise, we incorporate features that describe the statistics of luma, color, and gradient magnitude in our algorithm.

\subsubsection{Luma}
\begin{itemize}
    \item
    \textbf{NIQE naturalness}: We compute luminance features and a naturalness score based on them every $T'$ frames using the image naturalness index NIQE. They consist of parameters of GGD and AGGD fits to spatial MSCNs of selected patches of luma in each frame, and a naturalness score that measures the distance of the parameters from a statistical fit to a corpus of natural images. These features are averaged over the entire video over all non-overlapping groups of $T'$ frames.
    \item \textbf{$\sigma$ map:} We also model the statistics of the bandpass standard deviation $\sigma$ in (\ref{eq:sigma}). The $\sigma$ map is calculated using equation~\ref{eq:sigma}. The MSCNs of the $\sigma$ map also follows a GGD, as shown in~\cite{ruderman}, we extract the shape, variance, skewness, and kurtosis of the distribution. These features are averaged over the entire video clip. Previous research~\cite{videval,tlvqm,gmsd} has also shown that standard deviation pooling of features over time is useful for video quality assessment. We therefore compute the standard deviation of these features over every non-overlapping five-frame interval, and average the standard deviation values over the entire video. Again, $T'=5$ frames are used for ST chip computation, hence the pooling is copacetic with the computation of ST Chips. 
\end{itemize}

\subsubsection{Color}

We also model the statistics of the chrominance of videos. We use the CIELAB~\cite{cielab} color space, which is designed to model human perception of color. CIELAB has a luminance channel ($L*$) and two chrominance channels ($a^*$ and $b^*$), where $a*$ denotes the position of the color along the red-green axis, and $b*$ denotes the position of the color on the yellow-blue axis. Chroma ($C$) captures the intensity of a color, and is defined as 
\begin{equation}
    C = \sqrt{a^{*2}+b^{*2}}
\end{equation}

We compute the chroma map for each frame in the video, and find the MSCNs of the chroma map. The MSCNs are known to follow a first order GGD~\cite{friquee}, and the shape, variance, skewness, and kurtosis are extracted from the empirical distribution of each frame. 

We also compute the $\sigma$ map of the chroma, using (\ref{eq:sigma}). We extract the shape, variance, skewness, and kurtosis of the distribution. These features are averaged over the entire video clip. We also find the average standard deviation of features over $T'=5$ frame intervals, as described for the $\sigma$ calculations in luma space.

\subsection{Gradients}

Gradients are known to capture important information about edges, and since some distortions modify (reduce or increase) gradients, they have been effectively used to predict video quality in~\cite{friquee,higrade,tlvqm,videval,rapique}. We find the gradients of the luminance in the vertical and horizontal directions using a Sobel kernel of size 3. We then find the gradient magnitude at each pixel and compute the MSCNs of the gradient magnitude field. The second order statistics of the MSCNs of the gradient magnitude are particularly useful for predicting video quality, when combined with the features defined previously. The paired products of the MSCNs of the gradient magnitude are computed using (\ref{eq:pair}), and modelled using AGGDs (\ref{eq:aggd}). Four parameters are extracted on each paired product of the gradient magnitude at two scales. Just as for chroma, we average these features across time and also find the average standard deviation over groups of 5 frames.

\subsection{Quality Assessment}
Table~\ref{feat} gives a summary of all the features used in ChipQA. A total of 221 features are extracted on each video, starting from the $T'=5$\textsuperscript{th} frame. These are trained with a support vector regressor, as described in the following section.

\begin{table*}
\caption{Descriptions of features in ChipQA.}
\begin{center}
\begin{tabular}{|p{0.15\linewidth} | p{0.6\linewidth}|p{0.15\linewidth}|}
\hline
Domain &  Description & Feature index\\
\hline
Chroma & Average across all frames of GGD shape, GGD scale, skewness, and kurtosis at two scales. & $f_1-f_8$ \\
\hline
Chroma $\sigma$ map & Average across all frames of GGD shape, GGD scale, skewness, and kurtosis at two scales. & $f_9-f_{16}$ \\
\hline
Gradient &  Average across all frames of four parameters from AGGD fitted to pairwise products at two scales. & $f_{17}-f_{48}$ \\
\hline
Luma $\sigma$ map &  Average across all frames of GGD shape, GGD scale, skewness, and kurtosis at two scales. & $f_{49}-f_{56}$ \\
\hline
Chroma & Standard deviation across 5 frames of GGD shape, GGD scale, skewness, and kurtosis at two scales, averaged across all non-overlapping groups of 5 frames. & $f_{57}-f_{64}$ \\
\hline
Chroma $\sigma$ map &  Standard deviation across 5 frames of GGD shape, GGD scale, skewness, and kurtosis at two scales, averaged across all non-overlapping groups of 5 frames. & $f_{65}-f_{72}$ \\
\hline
Gradient &  Standard deviation across 5 frames of four parameters from AGGD fitted to pairwise products at two scales, averaged across all non-overlapping groups of 5 frames. & $f_{73}-f_{104}$ \\
\hline
Luma $\sigma$ map &  Standard deviation across 5 frames of GGD shape, GGD scale, skewness, and kurtosis at two scales, averaged across all non-overlapping groups of 5 frames. & $f_{105}-f_{112}$ \\
\hline
NIQE naturalness & Average over every 5 frames of features and scores of spatial naturalness index NIQE. & $f_{113}-f_{149}$ \\
\hline
ST-Chip &  Average over all chips of shape and scale parameters from GGD fits and four parameters from AGGD fits to pairwise products at two scales. & $f_{150}-f_{185}$ \\
\hline
ST Gradient Chips &  Average over all chips of shape and scale parameters from GGD fits and four parameters from AGGD fits to pairwise products at two scales. & $f_{186}-f_{221}$ \\
\hline
\end{tabular}
\label{feat}
\end{center}
\vspace{-10mm}
\end{table*}
\begin{figure*}[!h]
\centering
\subfloat[GGD fitted to ST Chip distribution]{{\includegraphics[width=0.45\textwidth]{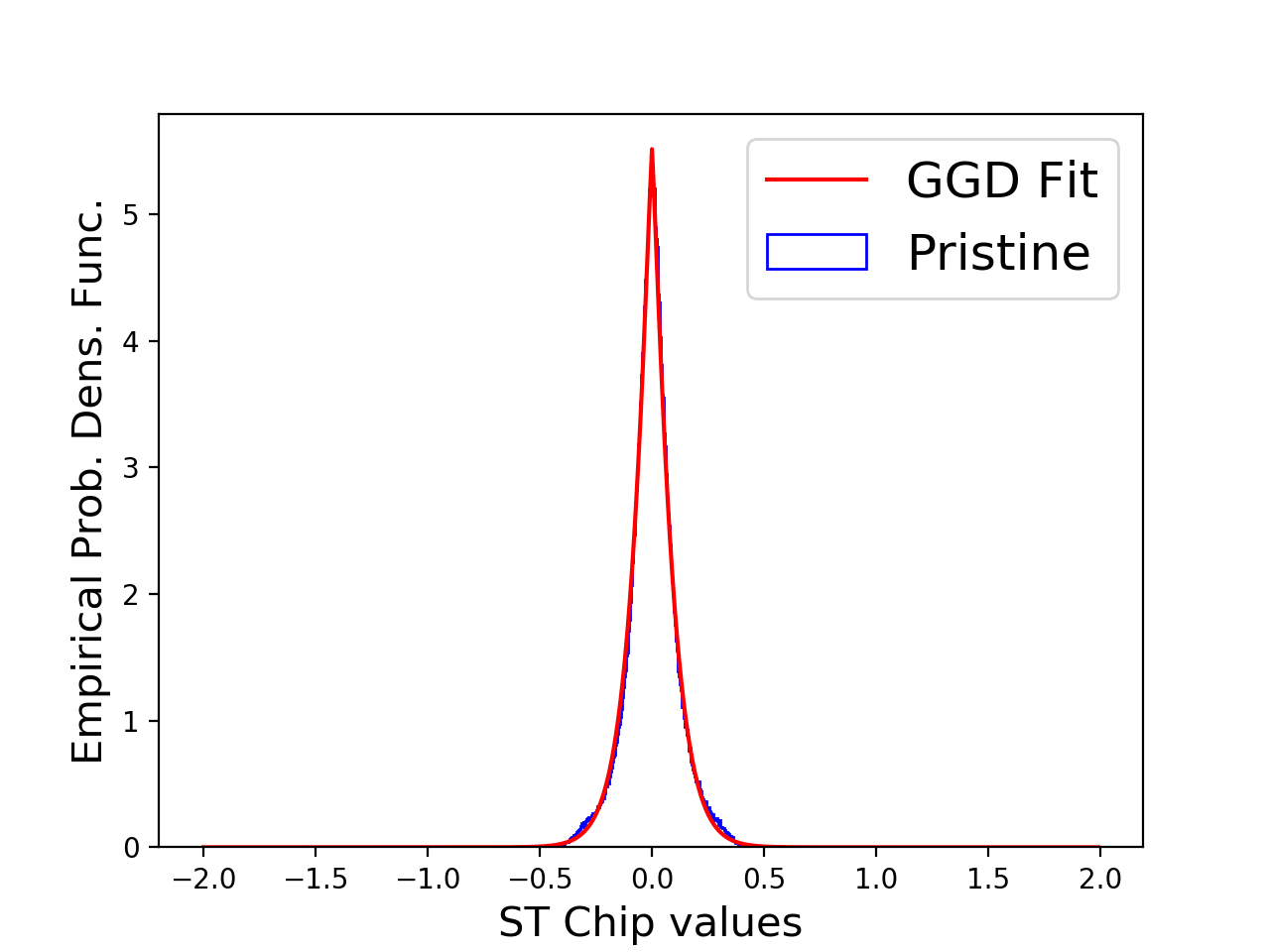}}}
\subfloat[GGD fitted to ST Gradient Chip distribution]{{\includegraphics[width=0.45\textwidth]{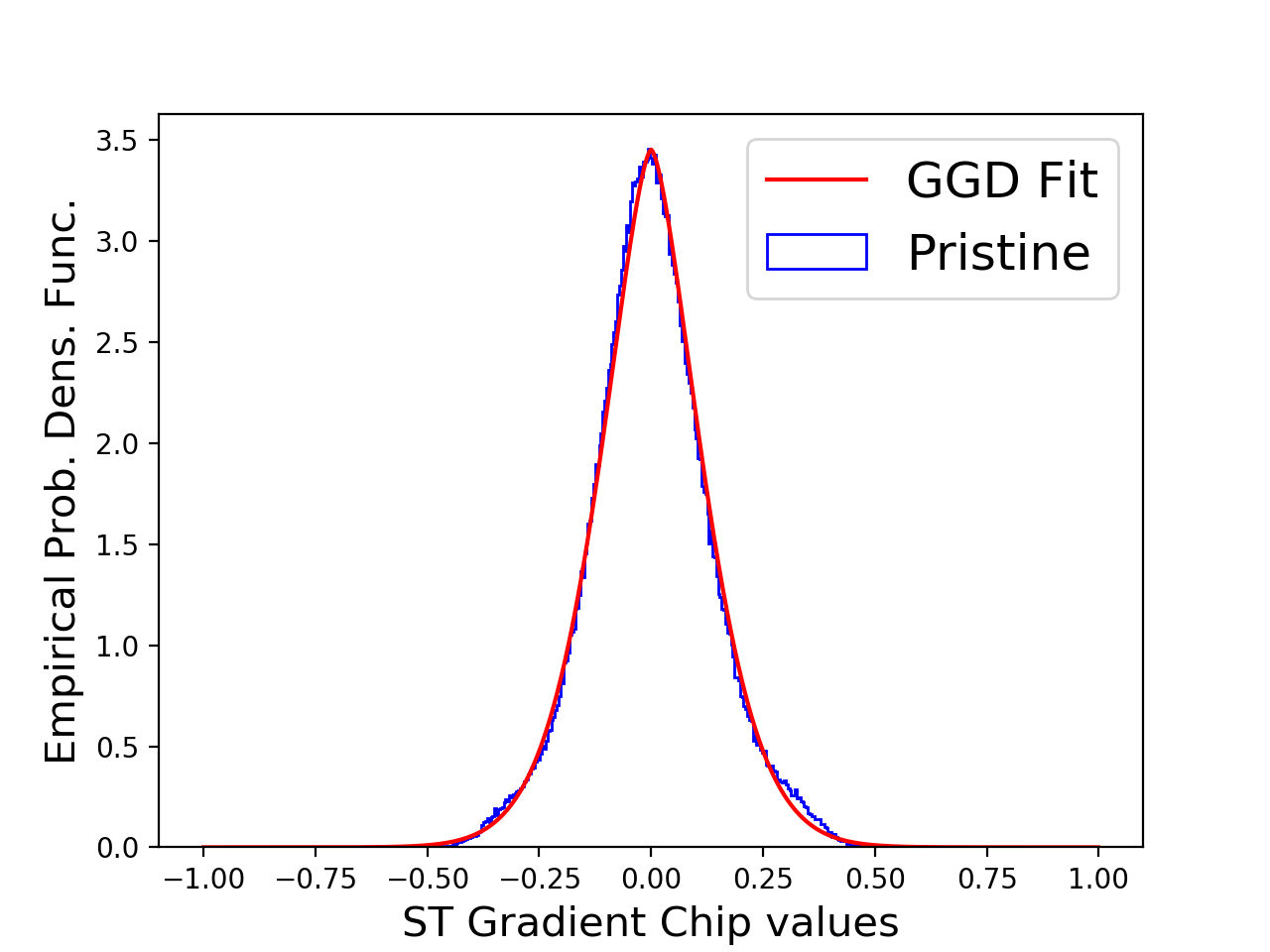}}} \\
\subfloat[AGGD fitted to ST Chip diagonal paired product distribution]{{\includegraphics[width=0.45\textwidth]{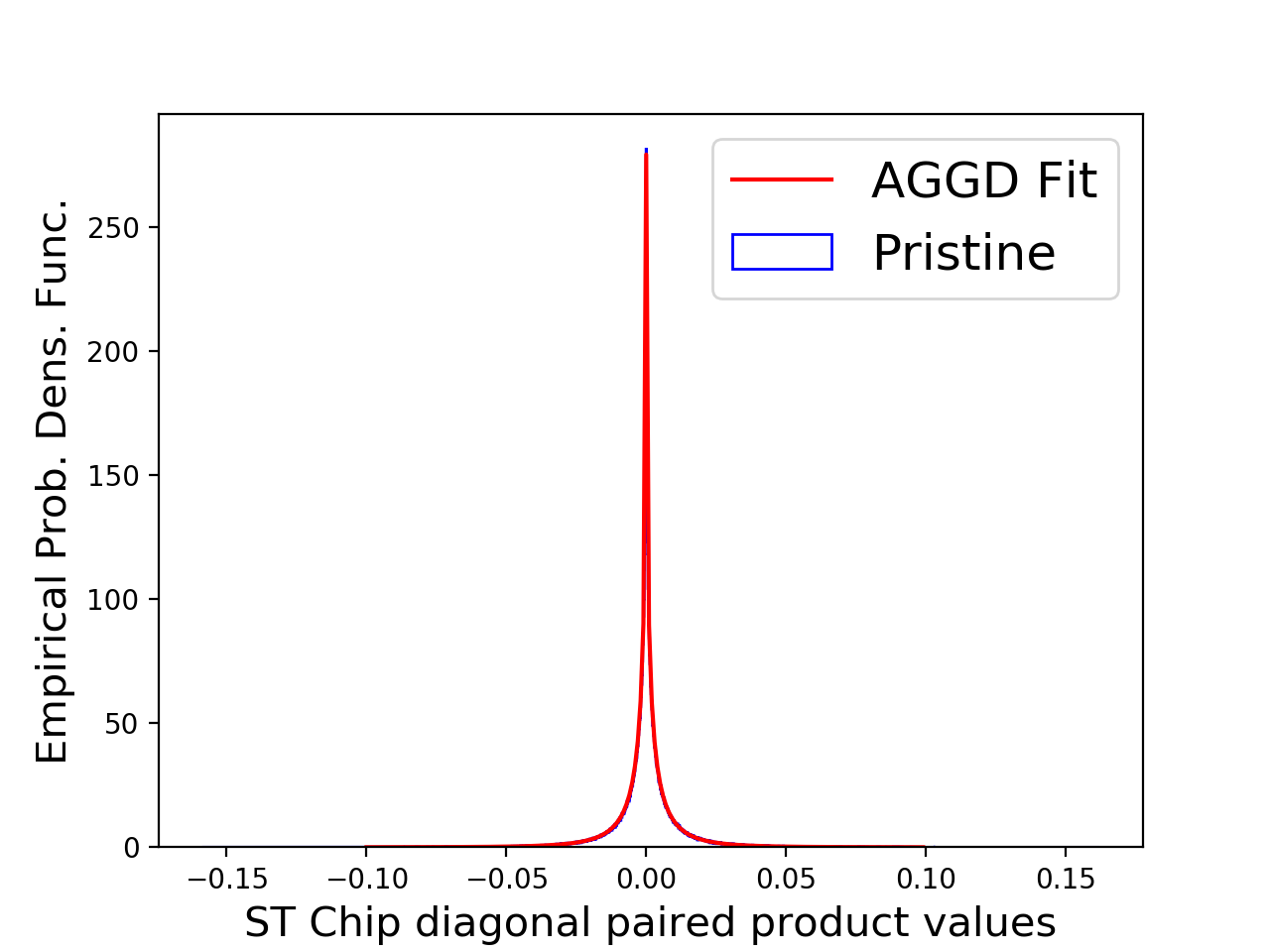}}}
\subfloat[AGGD fitted to ST Gradient Chip horizontal paired product distribution]{{\includegraphics[width=0.45\textwidth]{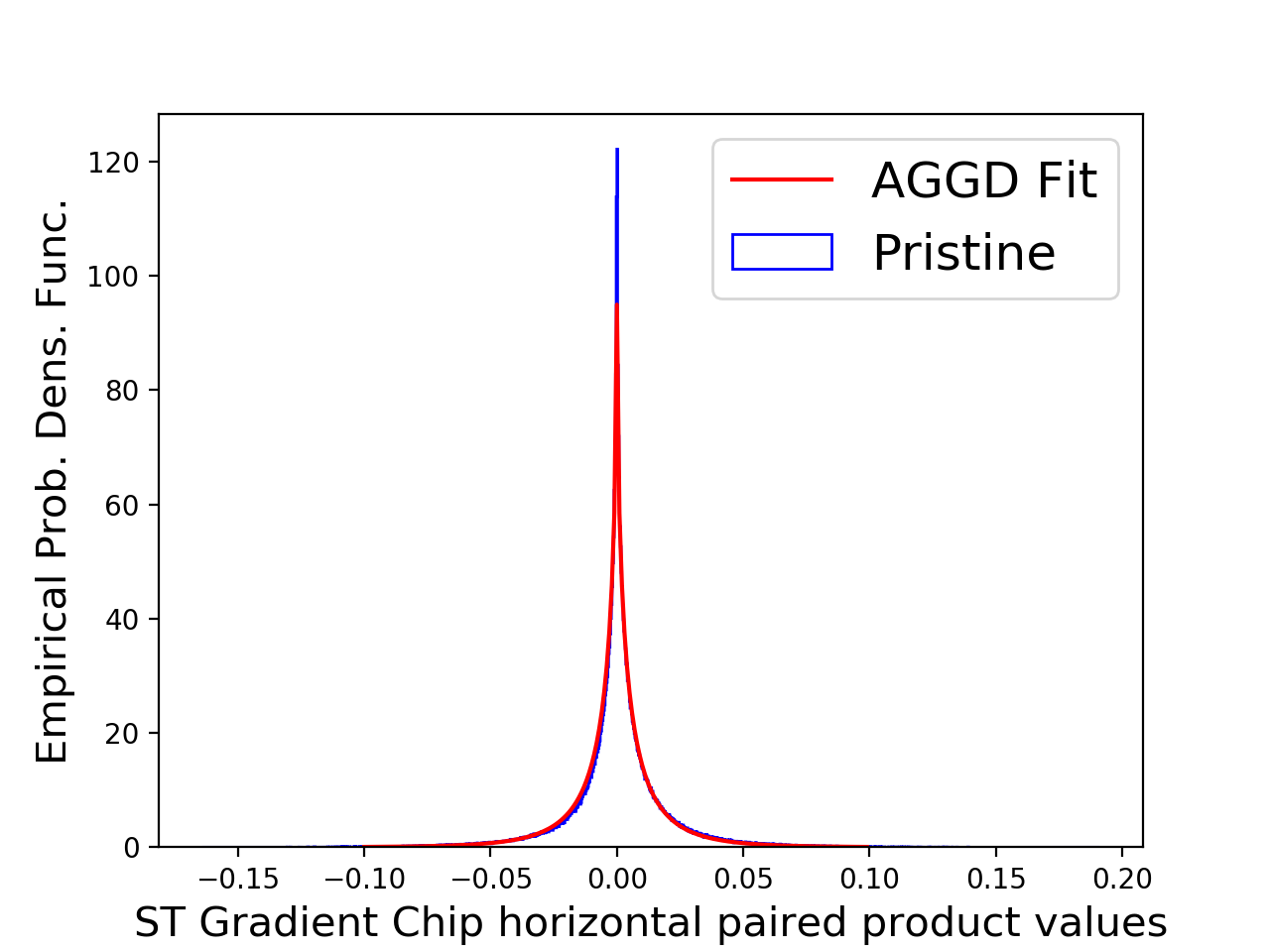}}}
\caption{Fits to empirical distributions. Fits are in red and histograms are in blue.}
\label{fig:fits}
\end{figure*}

\section{Experiments and Results}
\subsection{Validating ST Chips}
We conducted a series of experiments to examine the effectiveness of our method of generating ST chips. First, we studied how well our method is able to predict motion directions. The goal of our design is not to conduct very accurate motion estimation, since it is not necessary for the end goal of perceptual video quality measurement; the perception of absolute motion by the human visual system is not particularly accurate. Instead, we are interested in obtaining motion data that is consistent with statistical information available to human perception, rather than from complex search or optimization processes. Nevertheless, it is of interest to evaluate the efficacy of our simple kurtosis-based motion orientation selector. However, several challenges arise when attempting to quantitatively evaluate ChipQA's implicit motion orientation estimation process:
\begin{itemize}
    \item Many popular optical flow databases~\cite{middlebury,flyingchairs,kitti} only have optical flow on pairs of images, while ChipQA requires at least 5 frames to select motion directions.
    \item ChipQA only finds the directions of motions, not the magnitudes. Therefore some standard metrics such as endpoint error (EPE) cannot be used to evaluate ChipQA. 
    \item ChipQA computes the motion directions of patches, while most optical flow algorithms and ground truth data compute motion at the pixel level.
    \item ChipQA computes motion directions in quantized steps of $\pi/6$ from 0 to $\pi$, while the ground truth data in optical flow databases are generally more fine-grained.
    \item ChipQA does not differentiate between the angles $\theta$ and $\pi+\theta$ of any motion direction $\theta$, since the chips along both directions are the same.
\end{itemize}
The Sintel optical flow database~\cite{sintel} is a viable way to examine the implicit motion estimation of ChipQA, since it has optical flow ground truth on entire videos. Sintel is not a database of natural videos but of 3D animated films, so the videos from this database do not necessarily obey natural video statistics models. However, since animations are often made to be reasonably naturalistic, we have found that  natural video statistics models appear to hold well on this database. \par
As discussed earlier, absolute motion metrics are not appropriate for evaluating ChipQA. We therefore used the mean absolute angular deviation (MAAD) between the ground truth angular directions of motion $\theta$ (in radians) and the angles of motion $\theta'$ predicted by ChipQA:
\begin{equation}
    \text{MAAD}(\theta,\theta') = |\theta-\theta'|.
\end{equation}
We compared ChipQA's performance against the classical New Three Step Search algorithm (NTSS)~\cite{ntss}, which is a block motion algorithm, and the Farneb\"{a}ck dense optical flow predictor~\cite{farneback}. NTSS is used in VBLIINDS and some implementations of the MPEG and  H.26x codec families. Farneb\"{a}ck is used in ChipQA-0 and is the optical flow algorithm in the popular OpenCV library. Since ChipQA computes motion over patches, in order to compare the ground truth motion angles with those of ChipQA, the ground truth motion vectors were first averaged across 5x5 patches and across 5 frames in time. We then found those angles along which these vectors are oriented and mapped them to $[0,\pi)$ by replacing all angles $\theta > \pi$ with $\theta-\pi$. We did the same for the motion vectors predicted by NTSS and Farneb\"{a}ck, so that they could be compared to ground truth in the same way as ChipQA. The results are shown in Table~\ref{tab:motion}, which shows that ChipQA was competitive with respect to MAAD, outperforming NTSS but was not as accurate than Farneb\"{a}ck. The effectiveness of ChipQA is quite remarkable, given its extremely simple design. Indeed, we view these experiments as suggestive of the type of information that might be used by the visual brain to compute motion.

\begin{figure} 
  \includegraphics[width=\linewidth]{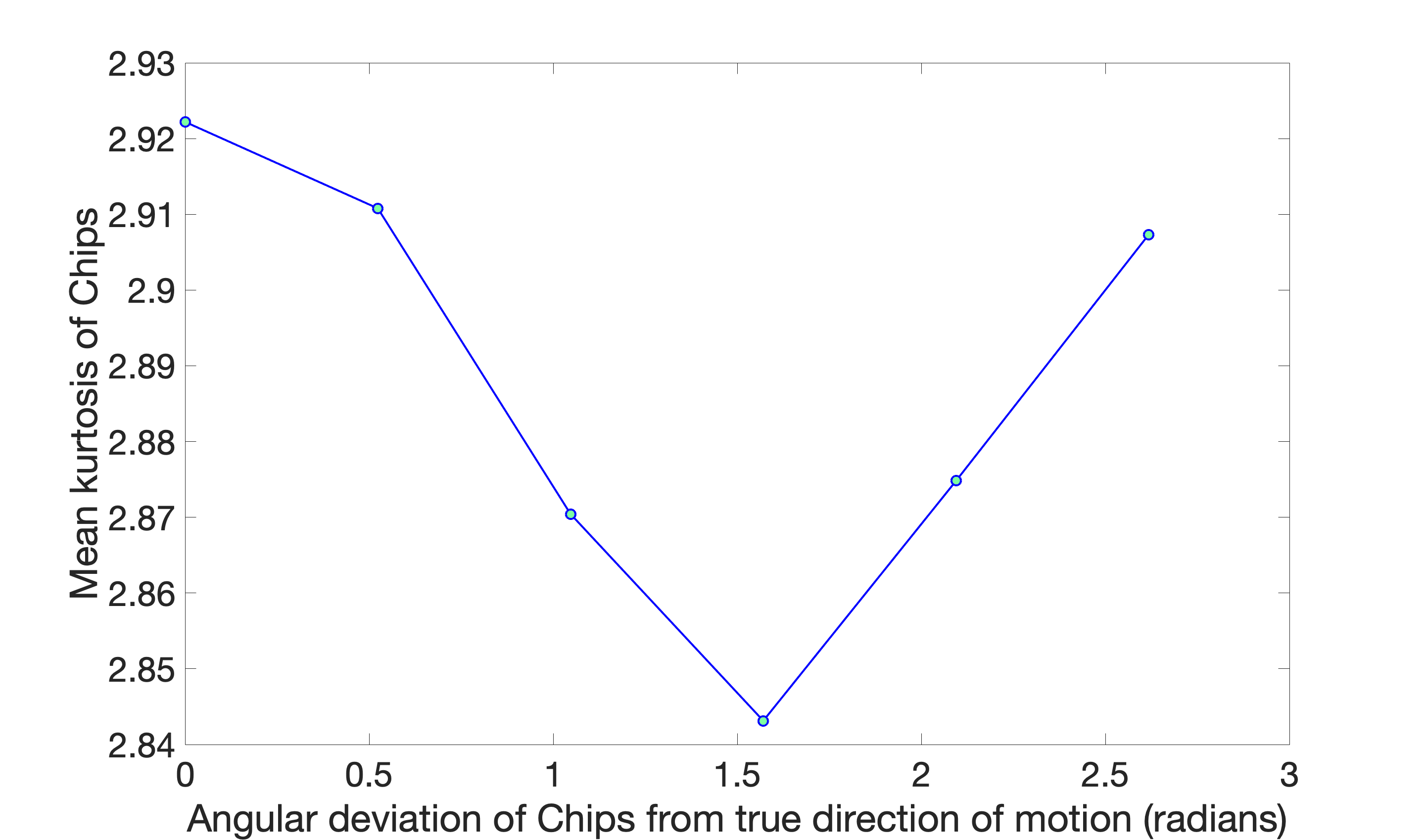}
  \caption{Mean kurtosis across all chips in Sintel vs. the angular deviation from the true motion direction.}\label{fig:kurt_vs_theta}
\end{figure}

We also conducted a separate experiment whereby we studied the variation of kurtosis of ST chips on the Sintel database, as the angular difference between the chip's orientation and the ground truth was varied. We computed the kurtosis of all the chips on the Sintel database (on all videos), for fixed differences between the true motion direction and each chip's orientation, and plotted the results as the differences were increased by steps of $\pi/6$, in Fig.~\ref{fig:kurt_vs_theta}.
When the angular difference was 0, i.e., when all the chips across all videos were exactly aligned with the ground truth motion vectors, the average kurtosis was closest to 3, the kurtosis of a Gaussian. As the angular difference was increased by steps of $\pi/6$, the average kurtosis diverged from 3 up to angular difference $\pi/2$, after which it again converged towards 3 as the angular cycle completed. This provides further validation of the simple angular motion estimator used in ChipQA, while providing insights into the relationship between natural video statistics and motion.

\begin{table}
\caption{Mean absolute angular difference (MAAD) between predicted and actual direction of motion $\theta$ for Sintel database}
\begin{center}
\begin{tabular}{|l|l|l|l|}
\hline
\textsc{Sequence} & ChipQA & \textsc{NTSS} & \textsc{Farneb\"{a}ck}\\
\hline
alley\_1 & 0.9133 \scriptsize(0.5975) & 1.2202 \scriptsize(0.5934) & 0.4295 \scriptsize(0.5410 )\\
\hline
alley\_2 & 1.0526 \scriptsize(0.7281)& 1.1773 \scriptsize(0.7723) & 0.7439 \scriptsize(0.8335 ) \\
\hline
ambush\_2 & 1.0453 \scriptsize(0.7261)& 0.9914 \scriptsize(0.6943)& 0.8535 \scriptsize(0.6754 ) \\
\hline
ambush\_4 & 1.0316 \scriptsize(0.7273) & 0.9871 \scriptsize(0.6842) & 0.9603 \scriptsize(0.7493 )\\
\hline
ambush\_5 & 0.9759 \scriptsize(0.6902)& 1.0201 \scriptsize(0.7237)& 0.9439 \scriptsize(0.7267 ) \\
\hline
ambush\_6 & 1.0354 \scriptsize(0.7381)& 0.8732 \scriptsize(0.5989)& 0.7790 \scriptsize(0.5633 ) \\
\hline
ambush\_7 & 1.0891 \scriptsize(0.7740) & 1.0856 \scriptsize(0.7536)& 0.9086 \scriptsize(0.8656 )\\
\hline
bamboo\_1 & 0.8173 \scriptsize(0.4868)& 0.9146 \scriptsize(0.4904) & 0.4114 \scriptsize(0.3262 )\\
\hline
bamboo\_2 & 1.3628 \scriptsize(0.8921)& 1.3069 \scriptsize(0.5563)& 1.5411 \scriptsize(1.3005 ) \\
\hline
bandage\_1 & 0.8503 \scriptsize(0.6008)& 0.7377 \scriptsize(0.5720)& 0.6774 \scriptsize(0.5623 ) \\
\hline
bandage\_2 & 0.9607 \scriptsize(0.6911)& 0.8218 \scriptsize(0.6342) & 0.8610 \scriptsize(0.7053 )\\
\hline
cave\_2 & 0.8766 \scriptsize(0.5860)& 0.8615 \scriptsize(0.5611) & 0.5518 \scriptsize(0.5526 )\\
\hline
cave\_4 & 0.9357 \scriptsize(0.6451) & 0.9018 \scriptsize(0.6236)& 0.6130 \scriptsize(0.6235 )\\
\hline
market\_2 & 0.8616 \scriptsize(0.5444)& 0.8612 \scriptsize(0.6297)& 0.3914 \scriptsize(0.4283 ) \\
\hline
market\_5 & 1.0810 \scriptsize(0.7546) & 1.0584 \scriptsize(0.7651)& 0.8849 \scriptsize(0.7895 )\\
\hline
market\_6 & 1.1404 \scriptsize(0.7934)& 1.1665 \scriptsize(0.7833) & 1.0946 \scriptsize(0.8966 )\\
\hline
mountain\_1 & 1.0020 \scriptsize(0.6989)& 1.0628 \scriptsize(0.7388)& 0.9432 \scriptsize(0.6799 ) \\
\hline
shaman\_2 & 0.8969 \scriptsize(0.6403)& 0.8540 \scriptsize(0.6044)& 0.6092 \scriptsize(0.6132 ) \\
\hline
shaman\_3 & 1.1720 \scriptsize(0.8081)& 1.2471 \scriptsize(0.8956)& 1.0625 \scriptsize(0.7017 ) \\
\hline
sleeping\_1 & 1.1975 \scriptsize(0.8345)& 1.2405 \scriptsize(0.8591)& 1.0812 \scriptsize(0.7469 ) \\
\hline
sleeping\_2 & 1.0476 \scriptsize(0.7201)& 1.0677 \scriptsize(0.7126)& 0.8090 \scriptsize(0.8862 ) \\
\hline
temple\_2 & 0.8724 \scriptsize(0.5483)& 0.8519 \scriptsize(0.5546)& 0.5240 \scriptsize(0.5110 ) \\
\hline
temple\_3 & 0.8945 \scriptsize(0.6144)& 0.9933 \scriptsize(0.6217)& 0.7312 \scriptsize(0.6574 ) \\
\hline 
\hline
ALL & 1.0048 \scriptsize(0.1294)& 1.0131 \scriptsize(0.1549) & 0.8002 \scriptsize(0.2607) \\
\hline
\end{tabular}
\label{tab:motion}
\end{center}
\vspace{-5mm}
\end{table}

\vspace{-3mm}
\subsection{Databases}
We evaluated our algorithm on four large databases, which are described below:
\begin{enumerate}
    \item LIVE Livestream VQA database~\cite{livestream} - 
    This is a new database containing 315 professional-grade videos with synthetic distortions. 45 videos are of pristine quality, and 7 different distortions were synthetically applied to each pristine video to create 315 synthetically distorted videos. The distortions that were applied are compression, aliasing, interlacing, judder, flicker, and frame-drop. All videos are of resolution 3840x2160, and the study was conducted on a 4K TV. 40 people participated in the study.
    \item LIVE ETRI database~\cite{etri} - The LIVE ETRI database contains 437 videos generated by applying various levels of combined space-time subsampling and video compression on 15 diverse video contents. The bitrates in this database vary from around 500 kbps to around 50 Mbps, and the database has 30 fps, 60 fps, and 120 fps videos. A total of 34 subjects took part in the study.
    \item YouTube UGC~\cite{youtube_ugc} - The YouTube UGC database consists of 1500 20 second clips sampled from millions of user-generated YouTube videos. Spatial and temporal features were used to ensure the sampled videos were diverse and representative of videos on YouTube. Each video was rated by more than 100 subjects. These videos are of different resolutions and include content from categories such as animation and VR.
    \item Konvid-1k~\cite{konvid} - This database consists of 1200 videos of authentically distorted user-generated content. All videos are of resolution 960x540. Videos in this database are known to not have significant temporal variation in quality~\cite{videval,pooling}. Each video has 114 subject ratings on average.
    \item LIVE Video Quality Challenge (VQC)~\cite{vqc} - LIVE VQC contains 585 videos of authentically distorted unique user-generated content. Each video was labeled by 40 human subjects on average. All videos are of user-generated content.
\end{enumerate}
\vspace{-3mm}
\subsection{Evaluation Details}

We used a support vector regressor (SVR) with a radial basis function as the learning engine for all our experiments. 20\% of the data was randomly selected for testing and 80\% was used for training and validation. 5-fold cross-validation was performed on the training set to find the best set of hyperparameters for the SVR. Content separation is performed for all experiments on the LIVE Livestream database and the LIVE ETRI database, hence videos containing the same content are always in the same fold of either training, validation, or testing. This prevents scores from being artificially boosted. We repeated 1000 random splits on LIVE Livestream and LIVE ETRI, which are databases containing synthetic distortions and reference videos. We conducted 100 splits for the Konvid-1K, VQC, and YT-UGC databases, which are UGC databases. Databases containing synthetic distortions generally result in larger standard deviations of the results, which is why we performed more train-validation-test splits on them. Grid search was performed over values of the kernel coefficient $\gamma$, and the regularization parameter $C$ that controls the squared L2 penalty. $\gamma$ was geometrically increased by 10 from $10^{-8}$ to $10$. $C$ was doubled from $2$ to $1024$. We report Spearman's Rank Order Correlation Coefficient (SROCC), Pearson's Linear Correlation Coefficient (LCC), and the root mean square error (RMSE) between the predicted scores and the mean opinion scores for each algorithm. SROCC measures the monotonicity of the relationship between the two quantities, while the LCC measures the linear correlation. Since the relationship between the predicted scores and the MOS may not necessarily be linear, the predicted scores $s$ were first passed through a logistic non-linearity~\cite{logistic}
\begin{equation}
f(s) = \beta_1(\frac{1}{2} - \frac{1}{(1+exp(\beta_2(s-\beta_3))})+\beta_4 x+\beta_5,
\end{equation}
before computing the LCC. The parameters are found by fitting $f(s)$ to the MOS.

\vspace{-3mm}
\subsection{Implementation Details}\label{exp_impl}
We used the luma definition from ITU Recommendation BT.709~\cite{bt709}. We use a frame-buffer of length $T'=5$ frames to compute ST Chips, keeping in mind potential applications of our algorithm in scenarios requiring low-latency, where using a larger buffer size might cause delays in the overall system. With $a=0.5$, the discrete 5-tap causal filter $k[n]$ is representative of the continuous filter $k(t)$. After $t=8$, there was a slight increase in the response which we rectified to 0 by choosing $P$ (length of filter) $= T’$ (length of frame buffer) $=5$, since in any case the filter converges to 0 as $t$ tends to infinity. Using a temporal filter of length greater than the frame buffer could cause serious boundary artifacts, hence we fixed the length of the filter ($P$) to be the length of the frame buffer ($T’$), which was fixed at 5. We also fixed $R=T'$, so that each ST chip is square and is chosen from a $5\times 5\times 5$ volume. The SROCC  values obtained on the LIVE Livestream VQA database for different values of $T'$ and $a$ are shown in Table~\ref{tab:atparameters}. The Table shows that taking $a=0.5$ and $R=5$ not only provides a discrete filter that is representative of $k(t)$, but also provides the best predictions. 

\begin{table}
\caption{Median SROCC for 1000 splits on the LIVE Livestream VQA database for different values of $K$, where window size is $2K+1 \times 2K+1$} 
\begin{center} 
\begin{tabular}{|l|l|l|l|l|}
\hline
$K$  & SROCC \\
\hline
3 &  0.7967\\
\hline
5 & 0.7882 \\
\hline
7 &  0.7963 \\
\hline
\end{tabular}
\label{tab:kl}
\end{center}
\vspace{-5mm}
\end{table}

\begin{table}
\caption{Median SROCC for 1000 splits on the LIVE Livestream VQA database for different values of $a$ and $R=T'$} 
\begin{center} 
\begin{tabular}{|l|l|l|l|l|}
\hline
$a$  & $R=T'$ & SROCC \\
\hline
0.25 & 5 & 0.7847 \\
\hline
0.25 & 9 & 0.7274 \\
\hline
0.5 & 5 & 0.7967 \\
\hline
0.75 & 5 & 0.7869  \\
\hline
0.75 & 9 & 0.7864 \\
\hline
\end{tabular}
\label{tab:atparameters}
\end{center}
\vspace{-5mm}
\end{table}
We used a window size of $2K+1\times 2K+1$, where $K=3$, for MSCN computation. We also studied how performance varied as this parameter was changed. Consistent with previous studies~\cite{brisque} of spatial MSCNs, we found that the size of the window did not significantly affect performance. We report the median SROCC over 1000 splits of the LIVE Livestream VQA database for different values of $K$, in Table~\ref{tab:kl}. Since increasing $K$ increases the computational cost, we used $K=3$ in our final implementation.  \par
A look-up table is used to implement the search for the best ST-chip direction. Coordinates along different directions from $\theta=0$ to $\theta = \pi/Q$ are pre-computed and rounded to the nearest integer, since pixel coordinates are integer values. The look-up table coordinates are computed using the polar form, and are indexed by the value of $\theta$ and $r$, where $r$ varies from $-(R+1)/2-1$ to $(R+1)/2$, where $R\times R$ is the dimension of each ST chip. Values are read from the look-up table during the search for the best direction. We fixed $Q=6$ in our implementation, and thus search $6$ directions at each $R\times R$ window to find the direction that best captures motion at that location. Results for different values of $Q$ are shown in Table~\ref{tab:qparam} for ChipQA for $R=5$. Increasing $Q$ improves performance, but beyond $Q=6$ performance seems to drop off.  \par
We also studied the effects of the chip downsampling factor $D$. The performance of the algorithm on the LIVE Livestream database for $D=1$, $D=4$ and $D=8$ are presented in Table~\ref{tab:downsample}. Increasing $D$ greatly reduces the cost of ChipQA (see Table~\ref{tab:compcost}), and using all of the chips is not necessary to conduct effective video quality assessment. We used $D=4$ in our final implementation.

\begin{table}
\caption{Median SROCC for 1000 splits on the LIVE Livestream VQA database for different values of chip downsampling factor $D$} 
\begin{center} 
\begin{tabular}{|l|l|l|l|l|}
\hline
$D$  & SROCC \\
\hline
1 &  0.7877\\
\hline
4 & 0.7967 \\
\hline
8 &  0.7920 \\
\hline
\end{tabular}
\label{tab:downsample}
\end{center}
\vspace{-5mm}
\end{table}

\subsection{Video Quality Assessment Results}\label{results}

\begin{figure*}[!h] 
  \includegraphics[width=\linewidth]{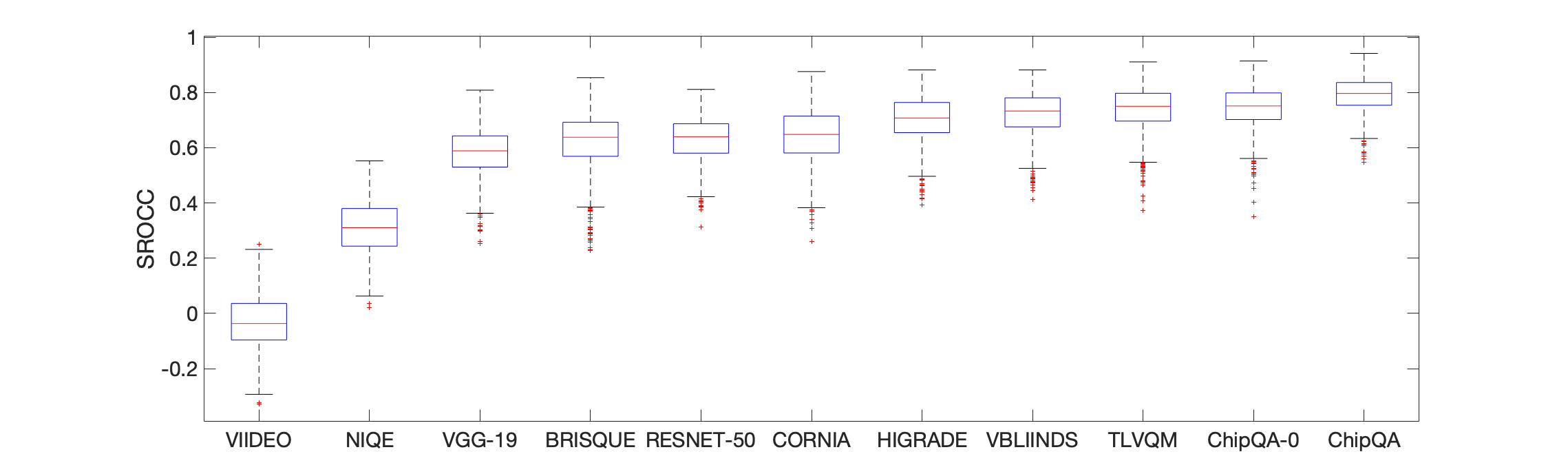}
  \caption{Box-plot of distribution of SROCCs for different algorithms for 1000 splits of the LIVE Livestream database.}\label{fig:boxplot}
\end{figure*}

Results on the LIVE Livestream VQA database, the LIVE ETRI database, Konvid-1k, LIVE VQC database and the  YouTube UGC  are shown in Tables~\ref{tab:liveapv},\ref{tab:etri}, \ref{tab:vqc}, \ref{tab:konvid}, and \ref{tab:ytugc} respectively. The scores of each  top-performing algorithm are boldfaced in each table. We also give box-plots of the SROCCs over the 1000 splits on the LIVE Livestream database in Figure~\ref{fig:boxplot}. ChipQA was able to achieve state-of-the-art performance on all databases. In particular, ChipQA significantly outperformed other compared algorithms on the LIVE Livestream database (Table~\ref{tab:liveapv}) and the LIVE ETRI database (Table~\ref{tab:etri}), both of which contain videos captured by professional videographers subjected to distortions commonly encountered at different stages of professional video capturing pipelines. These distortions include judder, frame-drop, space-time subsampling, etc. ChipQA produced the least variation in SROCC across the splits of the LIVE Livestream database as can be seen from the box plot and the standard deviations of the SROCCs reported in Table~\ref{tab:liveapv}.  ChipQA also performed very competitively on the UGC databases, which are known to be dominated by spatial distortions~\cite{pooling,videval}, hence image quality algorithms were among the top performing algorithms for UGC databases. 

We also evaluated two deep networks (VGG-19~\cite{vgg19} and ResNet-50~\cite{resnet}) on the LIVE Livestream database. Both networks were pre-trained on ImageNet and used to extract features from 25 random 227x227 crops from one frame per second on the LIVE Livestream database. The 4096 outputs of the fully connected layer of VGG-19, and the 2048 outputs of the average-pooled layer of ResNet-50 for each patch were averaged across all patches, forming  single 4096 and 2048 dimensional vectors, respectively, for each video. These video-level vectors were then used as features to train an SVR to predict video quality, using the same cross validation method as described earlier. These feature-extractors have been shown to perform quite well on full-reference tasks~\cite{lpips} as well as on UGC databases~\cite{videval}. However, they did not perform as well as the other methods on the LIVE Livestream database, which presents a significant variety and amount of temporal distortions. 

We ensured that videos of the same content did not appear in the same fold of training, validation, or testing. Reference videos and all of their distorted versions always appeared in the same fold. This ensured that the algorithm would not rely on learning  content quality, and to prevent overfitting. We also made use of the regularization parameter $C$, chosen by cross-validation to prevent overfitting by the SVR. We also measured the difference between the training and test SROCC to verify that ChipQA was not overfitting to the database, and found that the median difference between the SROCC on the training set and the SROCC on the test set was 0.1598 (16\%)

\begin{table}
\caption{Median SROCC for 1000 splits on the LIVE Livestream VQA database for different values of $Q$}
\begin{center}
\begin{tabular}{|l|l|}
\hline
$Q$  & SROCC \\
\hline
3  & 0.7852  \\
\hline
4  & 0.7658 \\
\hline
6  & 0.7967 \\
\hline
9 & 0.7626  \\
\hline
\end{tabular}
\label{tab:qparam}
\end{center}
\vspace{-5mm}
\end{table}

\begin{table*}
\caption{Median SROCC, LCC, and RMSE for LIVE Livestream Database. Standard deviations are in parentheses. Best performing algorithm is bold-faced.}
\begin{center}
\begin{tabular}{|c|c|c|c|c|}
\hline
\multicolumn{2}{|c|}{\textsc{Method}}  &  SROCC$\uparrow$ & LCC$\uparrow$ & RMSE$\downarrow$  \\  
\hline
\multirow{6}{*}{\textsc{Image Quality Metrics}} & NIQE\cite{niqe} & 0.3104 \scriptsize(0.0963) & 0.48674 \scriptsize(0.3323) & 11.2799 \scriptsize(1.0362) \\
\cline{2-5}
& VGG-19\cite{vgg19} & 0.5887 \scriptsize(0.0853) &  0.6600 \scriptsize(0.0687) &  9.3951 \scriptsize(0.6587) \\ 
\cline{2-5}
& ResNet-50\cite{resnet} & 0.6395 \scriptsize(0.0793) & 0.7011 \scriptsize(0.0666) & 8.9941 \scriptsize(0.6361) \\
\cline{2-5}
& BRISQUE\cite{brisque} &  0.6381 \scriptsize(0.1035) & 0.6841 \scriptsize(0.0835) & 9.1330 \scriptsize(0.9514)  \\
\cline{2-5}
& HIGRADE\cite{higrade} & 0.7088 \scriptsize(0.0805) & 0.7247 \scriptsize(0.0742) & 9.0958 \scriptsize(1.1249) \\
\cline{2-5}
& CORNIA\cite{cornia}  & 0.6482 \scriptsize(0.0977) & 0.6826 \scriptsize(0.0819) & 9.6472 \scriptsize(1.0725) \\
\hline
\multirow{4}{*}{\textsc{Video Quality Metrics}} &  VIIDEO\cite{viideo} & 0.0005 \scriptsize(0.0965) & 0.2643 \scriptsize(0.1670) & 12.0508 \scriptsize(0.9087) \\ 
\cline{2-5}
& V-BLIINDS\cite{vbliinds} & 0.7330 \scriptsize(0.0799) & 0.7609 \scriptsize(0.0699) & 8.2047 \scriptsize(1.0585) \\
\cline{2-5}
& TLVQM\cite{tlvqm} &  0.7503 \scriptsize(0.0815) & 0.7660 \scriptsize(0.0752) & 8.0392 \scriptsize(1.2181
)\\
\cline{2-5}
& ChipQA-0\cite{chipqa0} & 0.7513 \scriptsize(0.0762) & 0.7667 \scriptsize(0.0702) & 8.0195 \scriptsize(1.1147) \\
\cline{2-5}
& \textbf{ChipQA} & \textbf{0.7967 \scriptsize(0.0632)} & \textbf{0.8179 \scriptsize(0.0566)} & \textbf{7.6002 \scriptsize(1.0448)} \\
\hline
\end{tabular}
\label{tab:liveapv}
\end{center}
\vspace{-5mm}
\end{table*}

\begin{table*}
\caption{Median SROCC, LCC, and RMSE for LIVE ETRI. Standard deviations are in parentheses. Best performing algorithm is bold-faced.}
\begin{center}
\begin{tabular}{|c|c|c|c|c|}
\hline
\multicolumn{2}{|c|}{\textsc{Method}}  &  SROCC$\uparrow$ & LCC$\uparrow$ & RMSE$\downarrow$  \\  
\hline
\multirow{2}{*}{\textsc{Image Quality Metrics}} & NIQE\cite{niqe} & 0.3966 \scriptsize(0.1983) & 0.4435 \scriptsize(0.3049) & 12.2663 \scriptsize(1.6530) \\
\cline{2-5}
& BRISQUE\cite{brisque} &  0.2656 \scriptsize(0.3042) & 0.4315 \scriptsize(0.2461) & 12.3339 \scriptsize(1.5116)   \\
\hline
\multirow{4}{*}{\textsc{Video Quality Metrics}} & TLVQM\cite{tlvqm} &  0.2343 \scriptsize(0.1934) & 0.3018 \scriptsize(0.1957) & 13.0663 \scriptsize(1.2398)\\
\cline{2-5}
& V-BLIINDS\cite{vbliinds} & 0.4798 \scriptsize(0.1411) & 0.5344 \scriptsize(0.1194) & 11.7581 \scriptsize(1.4058) \\
\cline{2-5}
& ChipQA-0\cite{chipqa0} & 0.4012 \scriptsize(0.1591) & 0.4634 \scriptsize(0.1356) & 12.1476 \scriptsize(1.3464) \\
\cline{2-5}
& \textbf{ChipQA} & \textbf{0.6323 \scriptsize(0.1474)} & \textbf{0.6822 \scriptsize(0.1182)} & \textbf{10.0769 \scriptsize(1.4701)} \\
\hline
\end{tabular}
\label{tab:etri}
\end{center}
\vspace{-5mm}
\end{table*}
\begin{table*}
\caption{Median SROCC, LCC, and RMSE for Konvid 1k database. Standard deviations are in parentheses. Best performing algorithm is bold-faced.}
\begin{center}
\begin{tabular}{|c|c|c|c|c|}
\hline
\multicolumn{2}{|c|}{\textsc{Method}}  &  SROCC$\uparrow$ & LCC$\uparrow$ & RMSE$\downarrow$  \\  
\hline
\multirow{4}{*}{\textsc{Image Quality Metrics}} & NIQE\cite{niqe} & 0.3559 \scriptsize(0.0544) & 0.2533 \scriptsize(0.3338) & 0.6318 \scriptsize(0.0299) \\
\cline{2-5}
& BRISQUE\cite{brisque} (1 fps) &  0.5876{ \scriptsize(0.0422)} & 0.5989{ \scriptsize(0.0407)} & 0.5118{ \scriptsize(0.0244)}   \\
\cline{2-5}
& HIGRADE\cite{higrade} & 0.7310{ \scriptsize(0.0311)} & 0.7390{ \scriptsize(0.0296)} & 0.4306{ \scriptsize(0.0219)} \\
\cline{2-5}
& FRIQUEE\cite{friquee} (1 fps) & 0.7452{ \scriptsize(0.0274)} & 0.7506{ \scriptsize(0.0260)} & 0.4212{ \scriptsize(0.0208)} \\
\cline{2-5}
& CORNIA\cite{cornia} (1 fps) & 0.7685{ \scriptsize(0.0253)} & 0.7671{ \scriptsize(0.0240)} & 0.4093{ \scriptsize(0.0196)} \\
\hline
\multirow{4}{*}{\textsc{Video Quality Methods}} & VIIDEO\cite{viideo} & 0.3107{ \scriptsize(0.0492)} & 0.3261{ \scriptsize(0.0500)} & 0.6043{ \scriptsize(0.0243)} \\ 
\cline{2-5}
& \textbf{TLVQM}\cite{tlvqm} &  \textbf{0.7750{ \scriptsize(0.0242)}} & \textbf{0.7715{\scriptsize(0.0226)}} & \textbf{0.4058{ \scriptsize(0.0178)}}\\
\cline{2-5}
& V-BLIINDS\cite{vbliinds} & 0.7127{ \scriptsize(0.0355)} & 0.7085{ \scriptsize(0.0330)} & 0.4523{ \scriptsize(0.0210)} \\
\cline{2-5}
& ChipQA-0\cite{chipqa0} & 0.6973 \scriptsize(0.0311) & 0.6943 \scriptsize(0.0311) & 0.4600 \scriptsize(0.0221) \\
\cline{2-5}
& ChipQA & 0.7629 \scriptsize(0.0260) & 0.7625 \scriptsize(0.0256) & 0.4105 \scriptsize(0.0206) \\
\hline
\end{tabular}
\label{tab:konvid}
\end{center}
\vspace{-5mm}
\end{table*}
\begin{table*}
\caption{Median SROCC, LCC, and RMSE for LIVE VQC. Standard deviations are in parentheses. Best performing algorithm is bold-faced.}
\begin{center}
\begin{tabular}{|c|c|c|c|c|}
\hline
\multicolumn{2}{|c|}{\textsc{Method}}  &  SROCC$\uparrow$ & LCC$\uparrow$ & RMSE$\downarrow$  \\  
\hline
\multirow{4}{*}{\textsc{Image Quality Metrics}} & NIQE\cite{niqe} & 0.4603 \scriptsize(0.0735) & 0.3782 \scriptsize(0.4259) & 16.7678 \scriptsize(1.3756)\\
\cline{2-5}
& BRISQUE\cite{brisque} (1 fps) &  0.6192 \scriptsize(0.0529) & 0.6519 \scriptsize(0.0470) & 12.7489 \scriptsize(0.7664)   \\
\cline{2-5}
& HIGRADE\cite{higrade} & 0.6103 \scriptsize(0.068) & 0.6332 \scriptsize(0.065) & 13.027 \scriptsize(0.904) \\
\cline{2-5}
& FRIQUEE\cite{friquee} (1 fps) & 0.6579 \scriptsize(0.053) & 0.7000 \scriptsize(0.058) & 12.198 \scriptsize(0.914) \\
\cline{2-5}
& CORNIA\cite{cornia} (1 fps) & 0.6719{ \scriptsize(0.047)} & 0.7183{ \scriptsize(0.042)} & 11.832{ \scriptsize(0.700)} \\
\hline
\multirow{4}{*}{\textsc{Video Quality Metrics}} & \textbf{TLVQM}\cite{tlvqm} &  \textbf{0.8026 \scriptsize(0.0359)} & \textbf{0.7996 \scriptsize(0.0363)} & \textbf{10.1033 \scriptsize(0.7920)}\\
\cline{2-5}
& VIIDEO\cite{viideo} & -0.0336 \scriptsize(0.0770) & -0.0064 \scriptsize(0.0954) & 16.8163 \scriptsize(0.9094) \\ 
\cline{2-5}
& V-BLIINDS\cite{vbliinds} & 0.7005 \scriptsize(0.0457)& 7251 \scriptsize(0.0466) & 11.4744 \scriptsize(0.7483) \\
\cline{2-5}
& ChipQA-0\cite{chipqa0} & 0.6692 \scriptsize(0.0532) & 6965 \scriptsize(0.0477) & 12.1574 \scriptsize(0.8065) \\
\cline{2-5}
& ChipQA & 0.7192 \scriptsize(0.0550) & 0.7299 \scriptsize(0.0479) & 11.6014 \scriptsize(0.8370) \\
\hline
\end{tabular}
\label{tab:vqc}
\end{center}
\end{table*}

\begin{table*}
\caption{Median SROCC, LCC, and RMSE for YouTube UGC. Standard deviations are in parentheses.  Best performing algorithm is bold-faced.}
\begin{center}
\begin{tabular}{|c|c|c|c|c|}
\hline
\multicolumn{2}{|c|}{\textsc{Method}}  &  SROCC$\uparrow$ & LCC$\uparrow$ & RMSE$\downarrow$  \\  
\hline
\multirow{4}{*}{\textsc{Image Quality Metrics}} & NIQE\cite{niqe} & 0.1788 \scriptsize(0.050) & 0.1788 \scriptsize(0.094) & 0.6390 \scriptsize(0.026) \\
\cline{2-5}
& BRISQUE\cite{brisque} (1 fps) &  0.3820 \scriptsize(0.051) & 0.3952 \scriptsize(0.048) & 0.5919 \scriptsize(0.021)  \\
\cline{2-5}
& HIGRADE\cite{higrade} & 0.7376 \scriptsize(0.033) & 0.7216 \scriptsize(0.033) &  0.4471 \scriptsize(0.024)\\
\cline{2-5}
& \textbf{FRIQUEE}\cite{friquee} (1 fps) & \textbf{0.7652 \scriptsize(0.03)} & \textbf{0.7571 \scriptsize(0.032)}& \textbf{0.4169 \scriptsize(0.023)} \\
\cline{2-5}
& CORNIA\cite{cornia} (1 fps) & 0.5972{ \scriptsize(0.041)} & 0.6057{ \scriptsize(0.039)} & 0.5136{ \scriptsize(0.024)} \\
\hline
\multirow{4}{*}{\textsc{Video Quality Metrics}} & TLVQM\cite{tlvqm} &  0.6693 \scriptsize(0.03) & 0.6590 \scriptsize(0.03) & 0.4849 \scriptsize(0.022)\\
\cline{2-5}
& V-BLIINDS\cite{vbliinds} & 0.5590 \scriptsize(0.049) & 0.5551 \scriptsize(0.046) & 0.5356 \scriptsize(0.022) \\
\cline{2-5}
    & ChipQA-0\cite{chipqa0} & 0.5595\scriptsize(0.038)  & 0.5572 \scriptsize(0.040) & 0.5386 \scriptsize(0.024) \\
\cline{2-5}
& ChipQA & 0.7014 \scriptsize(0.031) & 0.6911 \scriptsize(0.033) & 0.4729 \scriptsize(0.025) \\
\hline
\end{tabular}
\label{tab:ytugc}
\end{center}
\end{table*}

\subsection{Performance by Distortion}

We also evaluated each algorithm on each single distortion in the LIVE Livestream database. The reference video and the synthetically distorted video for each content were used to evaluate the performance of each algorithm on each distortion separately. The results of this experiment are shown in Table~\ref{tab:sroccall} and Table~\ref{tab:plccall}. ChipQA performed very competitively on each distortion as well. The LIVE Livestream database has a number of scenes with high motion and ChipQA appears to be able to predict video quality well even for complex, high-motion scenes. TLVQM uses ``jerkiness" and ``jerkiness consistency" features which measure the similarity of motion vectors in the forward and backward directions, and which are  hand-crafted to capture frame drops, which is probably why it does so well on the frame drop category of distortions. CORNIA constructs codewords from image patches, which can capture simple patterns such as salt and pepper noise, blockiness, and interlacing. Interlacing has a very specific pattern of even and odd fields which lends itself to codeword construction, which is probably why CORNIA does so well on that category of distortions.

\begin{table*}[!h]
\renewcommand\arraystretch{1.3}
\centering
\caption{SROCC of the Compared NR VQA Models on different subsets of the LIVE Livestream database. The Scores of the Top Performing Algorithm in each category are Boldfaced}\label{tab:sroccall}
\begin{tabular}{|c|c|c|c|c|c|c|c|}
\hline
\textsc{Method} & \textsc{Overall} & \textsc{Compression} & \textsc{Aliasing} & \textsc{Judder} & \textsc{Flicker} & \textsc{Frame Drop} & \textsc{Interlacing}\\

\hline
NIQE(1 fps)& 0.3232 &0.2775  & 0.2860  & 0.2863  & 0.2832  &0.2842  & 0.2780 \\
\hline
BRISQUE(1 fps)&0.6381& 0.5748  & 0.7564   & 0.8235  & 0.6574   & 0.3849  &  0.8689 \\
\hline
VIIDEO & 0.0044 & 0.0053  & 0.0073  & 0.0055  & 0.0013  & 0.0024  & 0.0064  \\
\hline
CORNIA(1 fps)&0.6778 & 0.6894  & 0.7853  &0.7657   & 0.7049  & 0.2776  & \textbf{0.8824}  \\
\hline
HIGRADE(1 fps)& 0.6916 & 0.6244  & 0.7141  & 0.7441  & 0.6440  & 0.6130  & 0.8287  \\
\hline
V-BLIINDS& 0.7330 & 0.6450  &0.7606  & 0.8679  & 0.6182  & 0.7131  & 0.8060  \\
\hline
TLVQM & 0.7503 & 0.5614  & 0.7420  & 0.8328  & 0.6202  & \textbf{0.8555} &  0.8173  \\
\hline
ChipQA-0 & 0.7513 & 0.6594  & 0.7791 & 0.8513  & 0.6491  & 0.6780 &  0.8534  \\
\hline
ChipQA & \textbf{0.7575} & \textbf{0.7028}  & \textbf{0.7792}  & \textbf{0.8740}  & \textbf{0.7647}  & 0.7647 &  0.8513  \\
\hline
\end{tabular}
\end{table*}

\begin{table*}[!h]
\renewcommand\arraystretch{1.3}
\centering
\caption{PLCC of the Compared NR VQA Models on the LIVE Livestream database. The Scores of the Three Top Performing Algorithms Are Boldfaced.}\label{tab:plccall}
\begin{tabular}{|c|c|c|c|c|c|c|c|}
\hline
\textsc{Method} & \textsc{Overall} & \textsc{Compression} & \textsc{Aliasing} & \textsc{Judder} & \textsc{Flicker} & \textsc{Frame Drop} & \textsc{Interlacing}\\
\hline
NIQE(1 fps)& 0.4962 &0.2805 & 0.2865 & 0.2860  & 0.2848  & 0.2849  & 0.2850  \\
\hline
BRISQUE(1 fps)&0.6698& 0.7345  & 0.9321  & 0.8726  & 0.7268  & 0.3902   & {0.9118}  \\
\hline
VIIDEO & 0.2155 & 0.1222  & 0.1222  & 0.1235  & 0.1247  & 0.1256  & 0.1259  \\
\hline
CORNIA(1 fps)& 0.7257 & 0.8115  & {0.9508 } & 0.7875  &{ 0.8063 } &0.2138   & \textbf{{0.9142}}\\
\hline
HIGRADE(1 fps)&0.6990 &0.7595   & {0.9353}  & 0.7614  &0.6889   & 0.6062   & 0.8800   \\
\hline
V-BLIINDS& 0.7477 & {0.8055}  & 0.9202  &  {0.9200 }& 0.7086  & 0.7443   &  0.8873 \\
\hline
{TLVQM} & {0.7513} & 0.6788  & 0.9273   & {0.8914}  & {0.7724}  & \textbf{{0.8738}}  & 0.8358   \\
\hline
{ChipQA-0} & {0.7565} & {0.7783}  & {0.9490}   & 0.9071  & 0.6609  &{0.6945} & {0.9075}   \\
\hline
{ChipQA} &\textbf{0.7705} & \textbf{0.8719}  & \textbf{0.9735}   & \textbf{{0.9472}}  & \textbf{{0.8989}}  & {0.8730} & 0.9133   \\
\hline
\end{tabular}
\end{table*}

\subsection{Ablation Studies and Feature Ranking}

\begin{table}
\caption{Ablation study with median SROCC for 1000 splits on the LIVE Livestream VQA database as different feature spaces are removed}
\begin{center} 
\begin{tabular}{|c|c|c|}
\hline
\textsc{Rank} & \textsc{Feature Spaces Removed}  & SROCC \\
\hline
1 & ST chip+ST Gradient chip & 0.7376 \\ 
\hline
2 & Luma  & 0.7805 \\ 
\hline
3 & ST Gradient chip  & 0.7865 \\
\hline
4 & Gradient  & 0.7876 \\ 
\hline
5 & ST chip  & 0.7963 \\
\hline
6 & Chroma  & 0.7967 \\
\hline
7 & Chroma+Gradient & 0.8064 \\
\hline
\end{tabular}
\label{tab:ablapv}
\end{center}
\vspace{-5mm}
\end{table}

\begin{table}
\caption{Ablation study with median SROCC for 100 splits on the Konvid-1k database as different feature spaces are removed}
\begin{center} 
\begin{tabular}{|c|c|c|}
\hline
\textsc{Rank} & \textsc{Feature Spaces Removed}  & SROCC \\
\hline
1 & Chroma+Gradient & 0.7308 \\ 
\hline
2 & Luma  & 0.7448 \\ 
\hline
3 & Gradient  & 0.7460 \\
\hline
4 & ST chip + ST Gradient chip & 0.7552 \\ 
\hline
5 & Chroma  & 0.7589 \\
\hline
6 & ST Gradient chip  & 0.7641 \\
\hline
7 & ST chip  & 0.7669 \\
\hline
\end{tabular}
\label{tab:ablkvd}
\end{center}
\vspace{-5mm}
\end{table}

\begin{table*}
\caption{Feature ranking for the LIVE Livestream VQA database based on sequential forward selection}
\begin{center}
\begin{threeparttable}
\begin{tabular}{|p{0.1\linewidth}|p{0.1\linewidth}|p{0.6\linewidth}|p{0.1\linewidth}|}
\hline
\textsc{Rank} & \textsc{Feature index} & \textsc{Feature description}  & SROCC* \\
\hline
1 & 130 & Variance of right half of MSCN distribution of luma & 0.6515 \\ 
\hline
2 & 204 & GGD shape of ST chips at half-scale  & 0.7133 \\ 
\hline
3 & 149 & NIQE score  & 0.7411 \\
\hline
4 & 9 & GGD shape of MSCNs of Chroma $\sigma$ map & 0.7696 \\ 
\hline
5 & 25 & AGGD shape of pairwise products (along the main diagonal) of MSCNs of the gradient magnitude  & 0.8066 \\
\hline
6 & 150 &  GGD shape of ST chips at full-scale  & 0.8258 \\
\hline
7 & 161 & AGGD shape of pairwise products (along the main diagonal) of MSCNs of ST chips  & 0.8217 \\
\hline
8 & 123 & AGGD shape of pairwise products (along the main diagonal) of MSCNs of luma & 0.8043 \\
\hline
9 & 29 & AGGD shape of pairwise products (along the horizontal) of the gradient magnitude & 0.8062 \\
\hline
10 & 98 &  Variance of left half of MSCN distribution of pairwise product (along  the main diagonal) of MSCNs of the gradient magnitude at half-scale & 0.8134 \\
\hline
\end{tabular}
    \begin{tablenotes}
      \small
      \item *Note: This is the median SROCC obtained when the feature in the row entry is added to the set of the features ranked above it.
    \end{tablenotes}
 \end{threeparttable}
 \end{center}
\label{tab:featurerank}
\vspace{-5mm}
\end{table*}

We performed ablation studies of the studied feature spaces on the LIVE Livestream VQA database and the Konvid-1k database. The Konvid-1k database was chosen to be representative of UGC datasets. The results are shown in Table~\ref{tab:ablapv} and Table~\ref{tab:ablkvd}. Each feature space contains important quality-aware information about the video, but removing some of the spatial features (chroma and gradient features) appears to improve performance on LIVE Livestream. On the other hand, removing chroma and gradient features severely affects performance on Konvid-1k, showing that spatial features are vital for UGC but can only bring about the curse of dimensionality for datasets that have a significant amount of temporal distortion. We also find that removing ST chip features and ST gradient chip features from the algorithm severely affects performance on LIVE Livestream, but does not significantly affect performance on Konvid-1k. This suggests that temporal information is not as important in current UGC databases, validating earlier work that came to the same conclusion~\cite{videval,pooling}. 

We also studied performance when just the ST Chips, ST Gradient Chips, and the NIQE features were used. This set of features corresponds to what was used in ChipQA-0, but with our different and novel method of finding the chips, and removing all the other feature spaces that are in ChipQA but not in ChipQA-0. The median SROCC for just this set of features is 0.7901, greater than the 0.7513 SROCC for ChipQA-0, showing that ChipQA is conclusively better than ChipQA-0, and that our novel method of finding chips contributes significantly to the boost in performance.

We also ranked the importance of individual features in the database using sequential forward selection. We started with an empty set and then added a single feature to the feature set that maximized the median SROCC over 100 content-separated splits of the LIVE Livestream database. Repeating this process, we obtained an ordering of the features. The top 10 features and their descriptions are presented in Table~\ref{tab:featurerank}. We also report the median SROCC obtained when each feature is added to the set of features ranked higher than it. The NIQE score is, not expectedly, one of the top 10 features. The GGD shape of the ST chips at both scales are also among the top 10 features. {Gradients and chroma information also supply features that lie among the top 10.}

\subsection{Cross Database Performance}

We also evaluated the generalizability of ChipQA, TLVQM, and VBLIINDS across the studied databases. LIVE Livestream and LIVE ETRI are databases of professionally captured content and are thus treated separately from Konvid, VQC, and YT-UGC, which are UGC databases. Results of the cross-database performance comparisons on LIVE Livestream and LIVE ETRI are shown in Table~\ref{tab:crossdatabase_pro}, while results on the UGC databases are shown in  Table~\ref{tab:crossdatabase}. An SVR was trained on the features from the database in the row entry and the SROCC when the trained SVR was used to predict scores using features from the database on the column entry are reported. All of the compared methods yielded poor cross-database performance, primarily because these databases have very different characteristics in terms of both distortion type and content type. On UGC videos, cross database performance between Konvid and UGC was much better than on other databases among all the compared methods, probably because their contents are similar. YT-UGC spans a very wide range of content, including animations, gaming, and VR, hence these videos do not share many properties with the videos of natural scenes found in VQC and Konvid, which is probably why the cross-database performances were so poor on YT-UGC.  The distortions in the LIVE Livestream and LIVE ETRI databases are very different, involving high motions and temporal subsampling, respectively, though they both contain professionally-captured content. LIVE Livestream presents distortions such as interlacing, judder, flicker etc, while LIVE ETRI contains compression and temporal subsampling artifacts, which may explain the poor generalizability of all three VQA methods on these two databases.{In summary, NR VQA algorithms trained on one type of database sometimes perform quite poorly on others. While this might suggest weaknesses of the predictive models, more likely it is because the databases contain very different contents.}

\begin{table}
\caption{ Cross Database SROCC for methods trained on Row entry and evaluated on Column entry for databases with professionally-captured content.} 
\begin{center} 
\begin{tabular}{|l|l|l|l|l|}
\hline
 Method & & Livestream & ETRI \\
\hline
\multirow{2}{*}{\textsc{ChipQA}}  & Livestream &  - & 0.1286 \\
& ETRI & 0.2201 & -  \\
\hline
\multirow{2}{*}{\textsc{TLVQM}}  & Livestream &  - & 0.1478 \\
& ETRI & 0.2568 & - \\
\hline
\multirow{2}{*}{\textsc{VBLIINDS}}  & Livestream &  - &  0.3067 \\
& ETRI & 0.4331 & - \\
\hline
\end{tabular}
\label{tab:crossdatabase_pro}
\end{center}
\vspace{-5mm}
\end{table}

\begin{table}
\caption{ Cross Database SROCC for methods trained on Row entry and evaluated on Column entry for UGC databases.} 
\begin{center} 
\begin{tabular}{|l|l|l|l|l|l|}
\hline
 Method & & VQC & Konvid & YT-UGC \\
\hline
\multirow{3}{*}{\textsc{ChipQA}}  & VQC &  - & 0.5569 &  0.0757\\
& Konvid & 0.5478 & - & 0.0817\\
& YT-UGC & 0.5436  & 0.4030 & - \\
\hline
\multirow{3}{*}{\textsc{TLVQM}}  & VQC &  - & 0.6026 & 0.1768 \\
& Konvid & 0.6983 & - & 0.3241 \\
& YT-UGC & 0.3279  & 0.5588 & - \\
\hline
\multirow{3}{*}{\textsc{VBLIINDS}}  & VQC &  - & 0.6323 & 0.0969 \\
& Konvid & 0.6320 & - & 0.1677 \\
& YT-UGC & 0.1026  & 0.2072 & - \\
\hline
\end{tabular}
\label{tab:crossdatabase}
\end{center}
\vspace{-5mm}
\end{table}

\subsection{Computational Complexity}

We found the computational cost of each algorithm by computing the time each required to generate features on a single 4K video containing 210 frames. We also provide a crude estimate of the $O(n)$ computational complexity and the number of giga floating point operations (GFLOPS) for each algorithm. The results are shown in Table~\ref{tab:compcost}. ChipQA was the most efficient NR VQA algorithm, much more efficient than applying SOTA NR IQA algorithms on each frame. The algorithms are not implemented in the same language and may not be optimized, hence the computational times and FLOPS cannot be compared directly, but since most practitioners use these algorithms off-the-shelf, this provides a rough estimate of relative complexity. All the algorithms run in linear time but differ in the coefficient of linearity. The compute times for the IQA algorithms were obtained assuming they would be applied on all frames of the video. The times taken for FRIQUEE and HIGRADE are rough estimates obtained by finding the times taken on a single frame and multiplying by the number of frames. All the algorithms were run on an Intel Xeon E5-2620 CPU with a maximum frequency of 3 GHz. ChipQA is much faster than ChipQA-0, because it does not involve the use of optical flow, and also because it performs spatial and temporal downsampling.

\begin{table*}[!h]
\centering
\caption{Computational analysis of algorithms on a single 3840x2160 video with 210 frames from the LIVE Livestream VQA database}
\begin{threeparttable}
\begin{tabular}{|p{0.1\linewidth}|p{0.1\linewidth}|p{0.1\linewidth}|p{0.45\linewidth}|p{0.05\linewidth}|p{0.05\linewidth}|}
\hline
\textsc{Method}  & Number of features & Language & Computational Complexity & GFLOPS & Time (s) \\
\hline
BRISQUE\cite{brisque} & 36 & MATLAB & $\mathcal{O}(d^2NT)$ $d$: MSCN window size & 352 &  301 \\
\hline
ChipQA & 221 & Python & $\mathcal{O}((\frac{d^2}{D^2}+\frac{Q}{RD^2} + \frac{Q\log Q}{R^3D^2}) NT)$ $d$: MSCN window size, $D$: downsampling factor $Q$: chip search's quantization factor, $R$: size of each dimension of a chip  & 700 & 814  \\
\hline
TLVQM\cite{tlvqm} & 75 & MATLAB &  $\mathcal{O}((h_1^2N+k^2K)T_1+h_2^2NT_2)$ $h_1, h_2$: filter size, $k$: motion estimation block size, $K$: number of key points & 477 & 1002 \\
\hline
NIQE\cite{niqe} & 1 & MATLAB & $\mathcal{O}(d^2NT)$ $d$: MSCN window size & 3094 &  1008 \\
\hline
CORNIA\cite{cornia} & 10000 & MATLAB & $\mathcal{O}(d^2KNT)$,  $d$: window size, $K$: codebook size & 4480 & 2056\\
\hline
ChipQA-0\cite{chipqa0} & 109 & Python & $\mathcal{O}(d^2+1.5(1-0.5^{P+1})Iw^2)NT $,  $d$: MSCN window size, $P$: number of pyramids for optical flow, $I$: number of iterations per pyramid, $w$: window size for optical flow & 2605 & 2655  \\
\hline
VBLIINDS\cite{vbliinds} & 47 & Python & $\mathcal{O}((d^N+\log(k) N + k^2w^3)T)$, $d$: MSCN window size, $k$: DCT block size, $w$: motion vector tensor size & 465 &  3086 \\
\hline
VIIDEO\cite{viideo} & 1 & Python & $\mathcal{O}(N\log(N)T)$ & 2094 & 6403 \\
\hline
HIGRADE\cite{higrade} & 216 & MATLAB & $\mathcal{O}(3(2d^2+k)NT)$, $d$: MSCN window size, $k$: gradient kernel size & 9604 & 16240 \\
\hline
FRIQUEE\cite{friquee} & 560 & MATLAB & $\mathcal{O}((f^2dN+4N(\log (N) + m^2))T)$, $h_1,h_2$: filter size, $f$: number of color spaces, $k$: motion estimation block size, $K$: number of key points & 470,505 & 93240\\
\hline
\end{tabular}
    \begin{tablenotes}
      \small
      \item *Note: $N$ - number of pixels per frame. $T$ - number of frames in video. $T_1$ - total number of frames divided by 2. $T_2$ - number of frames sampled at 1 frame per second.
    \end{tablenotes}
 \end{threeparttable}
\label{tab:compcost}
\end{table*}

\section{Conclusion}
We presented a new algorithm for no-reference video quality assessment that is highly competitive with the state-of-the-art and is computationally very efficient. ST Chips are a novel feature space and the fact that they follow regular statistics represents a significant advance in our understanding of natural videos. We showed how these statistics can be parametrized, and how the parameters of the statistical fits can be used to predict video quality without the use of any distortion-specific features. ChipQA achieves high correlations with human judgments of video quality, especially of high motion videos, and is also computationally very efficient.


\section*{Acknowledgment}

The authors thank the Texas Advanced Computing Center (TACC) at The University of Texas at Austin for providing HPC resources that have contributed to the research results reported in this paper. URL: http://www.tacc.utexas.edu.

\ifCLASSOPTIONcaptionsoff
  \newpage
\fi



\bibliographystyle{IEEEtran}
\bibliography{chipqa_tip.bib}
%

%






\end{document}